\documentclass[aps,prb,twocolumn,superscriptaddress,showpacs,floatfix]{revtex4-1}

\usepackage{amsmath}
\usepackage{graphicx}
\usepackage{color}
\usepackage{epstopdf}
\usepackage{natbib}
\usepackage{xfrac}
\usepackage{wasysym}
\usepackage{multirow}

\usepackage{orcidlink}

\usepackage[caption=false]{subfig}

\usepackage{xcolor,colortbl}
\usepackage[normalem]{ulem}

\newcommand{\code}[1]{\texttt{#1}}

\definecolor{Gray}{gray}{0.85}
\definecolor{lblue}{rgb}{0.8,0.8,1}  

\definecolor{blau}{rgb}{0.,0.,1.}
\definecolor{gruen}{rgb}{0.,0.5,0.0}  \definecolor{orange}{rgb}{1,0.625,0.}  \definecolor{rot}{rgb}{1,0.0,0.}

\newcommand{\svo}{SrVO$_3$}
\newcommand{\sto}{SrTiO$_3$}

\newcommand{\pr}{^\prime}

\newcommand{\fref}[1]{Fig.~\ref{#1}}

\newcommand{\vek}[1]{ \hbox{\textbf #1}}

\renewcommand{\Im}{\hbox{Im}}
\renewcommand{\Re}{\hbox{Re}}

\newcommand{\out}[1]{}

\begin{document}
\title{
Particle-hole asymmetric lifetimes promoted by spin and orbital fluctuations\\
in SrVO$_3$ monolayers
}

\author{Matthias Pickem\,\orcidlink{0000-0002-0976-845X}}
\email{matthias.pickem@gmail.com}
\affiliation{Institute for Solid State Physics, TU Wien,  Vienna, Austria}
\author{Jan M.~Tomczak\,\orcidlink{0000-0003-1581-8799}}
\affiliation{Institute for Solid State Physics, TU Wien,  Vienna, Austria}
\author{Karsten Held\,\orcidlink{0000-0001-5984-8549}}
\affiliation{Institute for Solid State Physics, TU Wien,  Vienna, Austria}

\date{\today}

\begin{abstract}

The two-dimensional nature of engineered transition-metal ultra-thin oxide films offers a large playground of yet to be fully understood physics. 
Here, we 
study pristine SrVO$_3$ monolayers that have recently been 
predicted to display a variety of magnetic and orbital orders.
Above all ordering temperatures, we find that the associated non-local fluctuations lead to a momentum differentiation in the self-energy, particularly in the scattering rate. 
In the one-band 2D Hubbard model, momentum-selectivity on the Fermi surface ("$k=k_F$") is known to lead to pseudogap physics.
Here instead, in the multi-orbital case, we evidence a differentiation between momenta on the occupied ("$k<k_F$") and the unoccupied side ("$k>k_F$") of the Fermi surface. Our work, based on the dynamical vertex approximation, complements the understanding of spectral signatures of non-local fluctuations,  calls to (re)examine 
other ultra-thin oxide films and interfaces with methods beyond
dynamical mean-field theory, and may point to correlation-enhanced  thermoelectric effects.
\end{abstract}


\maketitle

\section{Introduction}

%
In the vicinity of phase transitions and in low-dimensional systems, non-local long-range
fluctuations are known to proliferate. These are not only crucial for the critical behavior
but may also lead to a strong enhancement of the scattering rate, i.e., a dampening of the quasiparticle lifetime.
 In three dimensions it is still debated \cite{Vilk1997,Rohringer2011,Rohringer2016}
 whether this scattering rate is actually diverging at a phase transition  or approaches a large but finite value.
 Even more peculiar is the situation in two dimensions. There, an actual phase transition---associated with the breaking of continuous symmetries---is prohibited. 
 Nonetheless non-local long-range fluctuation 
may still become huge  and can result in the famous pseudogap that has been experimentally observed in cuprate superconductors \cite{Norman1998,Timusk_1999,Norman2005,Keimer2015}. The pseudogap arises from a pronounced momentum differentiation of the scattering rate at low energy.
It is largest in the anti-nodal direction where eventually a gap opens at low enough temperatures. One possible explanation are long-range {\it antiferromagnetic} spin fluctuations \cite{PhysRevB.42.7967,Vilk_1996,Rost2012,PhysRevLett.93.147004,Gull2013,Cyr-Choiniere18},
with the momentum differentiation originating from the perfect antiferromagnetic nesting at the hot spots \cite{Kampf90,Pines93,Abanov03,Vilk1997,Vilk97-2,Wu20}, from the vicinity to a van Hove singularity in the anti-nodal direction \cite{Gonzalez00,Halboth20b,Honerkamp01,PhysRevResearch.2.033067},  or from the spin-fermion vertex turning complex at strong coupling \cite{Krien2021}.
It has also been suggested, on the basis of model studies, that a pseudogap phase can be driven by {\it ferromagnetic} fluctuations
\cite{PhysRevB.37.3299,PhysRevB.68.064408,PhysRevB.68.214405,PhysRevB.71.085105,PhysRevB.72.035111}.
\begin{figure*}[t!]
  \centering
  \subfloat[SrO-terminated monolayer]{
    \includegraphics[width=0.46\textwidth]{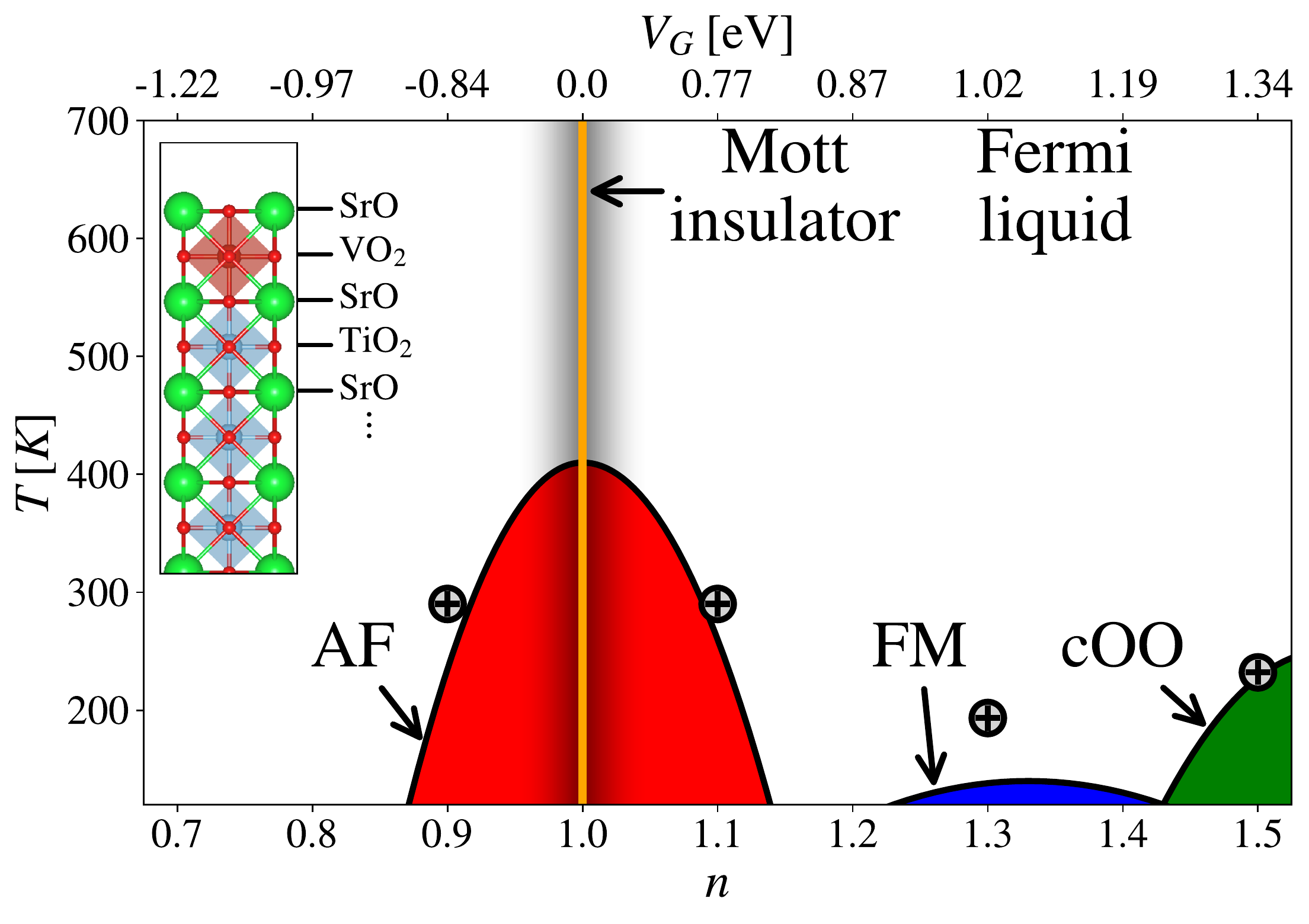}
    }
  \quad
  \subfloat[VO$_2$-terminated monolayer]{
    \includegraphics[width=0.46\textwidth]{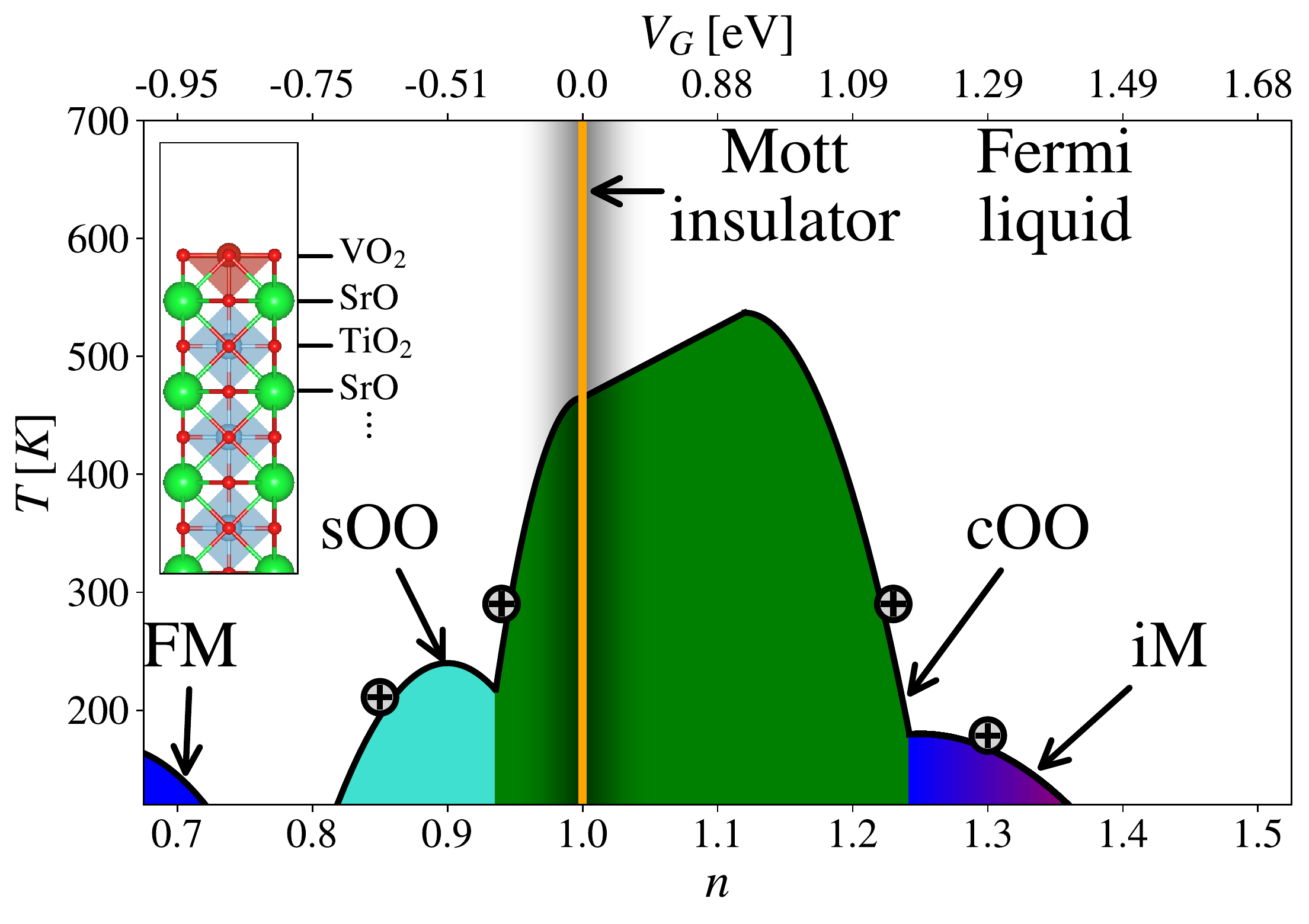}
    }
\caption{{\bf Phase diagrams.} (a) SrO- and (b) VO$_2$-terminated \svo-monolayer on top of a \sto\ substrate (see insets for crystal structures) exhibit numerous phases as a function of electrons per site ($n$; lower $x$-axis) in the low-energy vanadium $t_{2g}$ orbitals or gate voltage ($V_G$; upper $x$-axis):
    antiferromagnetism (AF: red), ferromagnetism (FM: blue), incommensurate magnetism (iM: blueish), checkerboard orbital order (cOO: green), stripe orbital order (sOO: turquoise). 
    The colored domes mark the occurrence of long-range order 
    within dynamical mean-field theory (DMFT);
    adapted from Ref.~\onlinecite{Pickem2021}.
   The ``$+$''--marks indicate points for which we present D$\Gamma$A (and DMFT) data in the present paper.   
    \label{fig:phasediagram}}
\end{figure*}
While pseudogap physics is mostly associated with cuprates, it has also been evidenced 
in other layered materials\cite{Benjamin_2D3D}: iron pnictides\cite{Xu2011,PhysRevLett.109.027006,PhysRevLett.109.037002,PhysRevB.89.045101} and chalcogenides\cite{PhysRevLett.111.217002},
(layered) nickelates\cite{PhysRevLett.106.027001}, and iridates\cite{Kim187}.
However, the origin of momentum-differentiated scattering rates is far from understood.

Here, we study a material that is two-dimensional by engineering:
 a monolayer of \svo. In the bulk, \svo\ is a correlated paramagnetic metal \cite{PhysRevB.58.4372,Mo2003} with a correlation-induced kink \cite{Nekrasov2006,Byczuk2007} linked to the 
 effective Kondo temperature \cite{Held2013}. 
Grown as a film, it is known to undergo a metal-insulator transition below
a critical thickness when deposited on an \sto\  \cite{PhysRevLett.104.147601,Kobayashi2017,PhysRevLett.114.246401}
or LSAT \cite{AMI.2014} substrate.
We focus on a  monolayer of \svo\ grown on the common  \sto\ substrate and consider two different terminations of the film to the vacuum:  VO$_2$ and SrO, see insets of \fref{fig:phasediagram}. Only the former has been evidenced experimentally  \cite{PhysRevLett.119.086801}, but it could be preferable to cover the films with a SrTiO$_3$ capping layer which leads to a structure more akin to the SrO termination. Such a capping layer also prevents a surface reconstruction with oxygen adatoms, which result in a dead surface layer \cite{Gabel2022}, at least for slightly thicker films. 
A preceding study \cite{Pickem2021}, based on  dynamical mean-field theory  (DMFT) \cite{PhysRevLett.62.324,PhysRevB.45.6479,Georges96}, revealed a  rich variety of orbital ordered and magnetic phases as a function of doping, see \fref{fig:phasediagram}. Experimentally, the phase diagrams could be perused   by applying a gate voltage.

In the present paper, we go beyond DMFT and study the effect of non-local fluctuations on spectral properties 
using the dynamical vertex approximation (D$\Gamma$A) \cite{toschi:045118,Katanin2009,Anna_ADGA,RevModPhys.90.025003}.
As for cuprates, we find that strong long-range fluctuations lead to a substantial momentum-dependence in the self-energy. 
In stark contrast to the cuprates, however, in ultrathin films of \svo\ 
the momentum differentiation does not distinguish momenta {\it on} the Fermi surface but those {\it perpendicular} to it. For example for the SrO-termination and antiferromagnetic spin fluctuations (above the red dome in \fref{fig:phasediagram}a), occupied states with momenta below the Fermi surface have a long, Fermi liquid-like lifetime. Instead, unoccupied states above the Fermi surface have short lifetimes and even kinks (downturns) in the self-energy, signaling a depletion of states. For orbital-ordering and ferromagnetic fluctuations (above green and blue regions) it is {\it vice versa}. 
The same is true for the VO$_2$-termination and the most relevant $xz$/$yz$ orbitals in the regime of orbital and incommensurate magnetic fluctuations above half-filling ($n>1$ in \fref{fig:phasediagram}b, above the green and blue regions). Below half-filling, instead, non-local correlations only have a minor impact on the self-energy of this termination.

A pronounced particle-hole asymmetry in the {\it real part} of the self-energy is a common phenomenon, mostly owing to non-local exchange. For example, in $GW$ calculations \cite{Godby88,PhysRevB.87.115110,Tomczak14} it leads to larger semiconductor band gaps than in density functional theory. For the {\it imaginary part} of the self-energy (the scattering rate), however, such an asymmetry has, to the best of our knowledge, not been reported so far.
The particle-hole asymmetric lifetimes arising here from a momentum-selectivity of renormalizations can complement more ubiquitous orbital-selective
asymmetries and could drive large thermoelectric effects.

The outline of the paper is as follows: Section
\ref{Sec:Method} provides information on the employed electronic structure methods.
Section \ref{Sec:DMFTspectra} presents the DMFT Fermi surfaces and spectral functions and the trends upon doping the SrVO$_3$ monolayer.
Non-local fluctuations in AbinitioD$\Gamma$A suppress ordering instabilities but strong long-range fluctuations persist and affect spectra and self-energies. An analysis of the AbinitioD$\Gamma$A results and the evidenced momentum selectivity are presented in Section \ref{Sec:ADGA},
and further discussed in Section \ref{Sec:discussion}.
Finally, Section
\ref{Sec:Conclusion} summarizes our conclusions.

\section{Method}%
\label{Sec:Method}

The crystal structures used are identical to Ref.~\onlinecite{Pickem2021}: 
Density function theory (DFT) calculations are based on the \code{WIEN2K} package \cite{wien2k,doi:10.1063/1.5143061} with PBE\cite{PhysRevLett.77.3865} as exchange-correlation potential.
We construct a slab as displayed in the insets of \fref{fig:phasediagram} consisting of  one unit cell of
\svo\ on top  six unit cells of the \sto\ substrate and surrounded (in $z$-direction) by sufficient vacuum of about $10$\AA\ to both sides.
While the in-plane lattice constant  of the heterostructure is locked to the (theoretical) \sto\ substrate\cite{PhysRevMaterials.3.115001} ($a^{\mathrm{PBE}}_{\mathrm{SrTiO}_3}=3.95$\AA)
all other internal atomic positions are relaxed, except for the two unit cells of \sto\ furthest away from \svo.
The \code{WIEN2K} band-structure is then projected
onto maximally localized V-t$_{\mathrm{2g}}$ Wannier orbitals, using the \code{WIEN2Wannier} \cite{wien2wannier} interface to \code{Wannier90} \cite{wannier90}.
The thus obtained Wannier Hamiltonian is supplemented by a Kanamori
interaction using $U=5$eV, $J=0.75$eV, $U'=3.5$eV,
and solved
by dynamical mean-field theory (DMFT) \cite{Georges96,doi:10.1080/00018730701619647}. Doping is modeled by a posterior-to-DFT
adjustment of the chemical potential in DMFT.
For DMFT spectral functions, analytic continuation was performed with the maximum entropy method implemented in \code{ana\_cont} \cite{ana_cont,kaufmann2021anacont}. There, the hyper parameter $\alpha$ was determined with the chi2kink method and a preblur window size of $\sigma=0.05$eV was employed.

In this paper we go beyond DMFT \cite{Tomczak2017review,RevModPhys.90.025003} and treat non-local correlations  in the \svo\ film with \code{AbinitioD$\Gamma$A} \cite{Anna_ADGA, JPSJ_ADGA, CPC_ADGA}. Contrary to finite-size cluster methods,  the D$\Gamma$A approach\cite{toschi:045118,Katanin2009,RevModPhys.90.025003}
and other, closely related diagrammatic extensions of DMFT \cite{Kusunose2006,PhysRevB.77.033101,Rohringer2013,Taranto2014,Li2015,TRIQS,RevModPhys.90.025003}
are not limited to short-range fluctuations. It well describes pseudogaps induced by antiferromagnetic fluctuations in the one-band 2D Hubbard model \cite{Gull2013,Sordi2011,Sakai12,Schaefer2015-3,Schaefer2015-2,Gunnarsson2015,Gukelberger2016,Schaefer2020} and (quantum) critical behavior \cite{Rohringer2011,Antipov2014,Schaefer2017,Schaefer2019}.
Orbital-ordering and ferromagnetic fluctuations have hitherto not been studied by D$\Gamma$A or other diagrammatic extensions of DMFT.
For the  \code{AbinitioD$\Gamma$A}, we here calculate the local particle-hole irreducible vertex at DMFT self-consistency  by continuous-time quantum Monte Carlo simulations in the hybridization expansion \cite{Werner2006,Gull2011a} using \code{w2dynamics} \cite{w2dynamics} with worm sampling \cite{Gunacker15}. From this  local vertex we subsequently calculate the particle-hole and transversal particle-hole Bethe-Salpeter ladder diagrams, and, through the Schwinger-Dyson equation, the non-local self-energy. This way we include non-local correlation effects in the self-energy. The D$\Gamma$A chemical potential was readjusted to fix the total number of electrons to the considered doping level.
In this study we apply D$\Gamma$A in a one-shot setting, forgoing non-local self-consistency \cite{PhysRevB.103.035120}.
For a review of the method, see Ref.~\onlinecite{RevModPhys.90.025003}; for computational details of the  AbinitioD$\Gamma$A see Ref. \onlinecite{CPC_ADGA}.
D$\Gamma$A and DMFT Fermi surfaces were obtained from the  Green's function at imaginary time $\tau=\beta/2$ ($\beta=1/k_B T$). This procedure corresponds to a spectral function $\overline{A(\mathbf{k},\omega=0)}$ that is averaged over a frequency-interval $\sim k_B T$ around the Fermi level.


\section{DMFT: orbital effects}
\label{Sec:DMFTspectra}
The DMFT phase diagram  \fref{fig:phasediagram} shows, as a function of doping and surface termination, a rich variety of different magnetic and orbitally ordered phases \cite{Pickem2021}.
Non-local fluctuations will strongly suppress the DMFT phase transitions in two dimensions but lead, at the same time, to strong scattering rates and self-energy corrections. In Sec.~\ref{Sec:ADGA}, we will study these renormalizations using the D$\Gamma$A at the $(n,T)$-points indicated in \fref{fig:phasediagram}. The temperatures have been chosen so that we are close to the respective phase transitions in DMFT and, thus, can expect pronounced non-local correlations%
\footnote{Due to the multi-orbital nature and the temperature scaling of the D$\Gamma$A we are restricted in temperature, i.e. we are not able to approach the FM instability much further.}.
Before turning to these D$\Gamma$A results, in this Section  we first analyze the  DMFT Fermi surfaces and ${\mathbf k}$-integrated spectra at the same fillings (orbitally resolved occupations, DMFT susceptibilities, and  ${\mathbf k}$-integrated DMFT spectra at filling $n=1$ with and without crystal field splitting have already been presented in Ref.~\onlinecite{Pickem2021}). In DMFT, non-local fluctuations are not included and thus do not affect the self-energy and spectral function. 
As a consequence, approaching the ordered states does not result in a pronounced temperature dependence of the DMFT spectra and self-energy.

\begin{figure*}[tb]
  \includegraphics[width=1\textwidth]{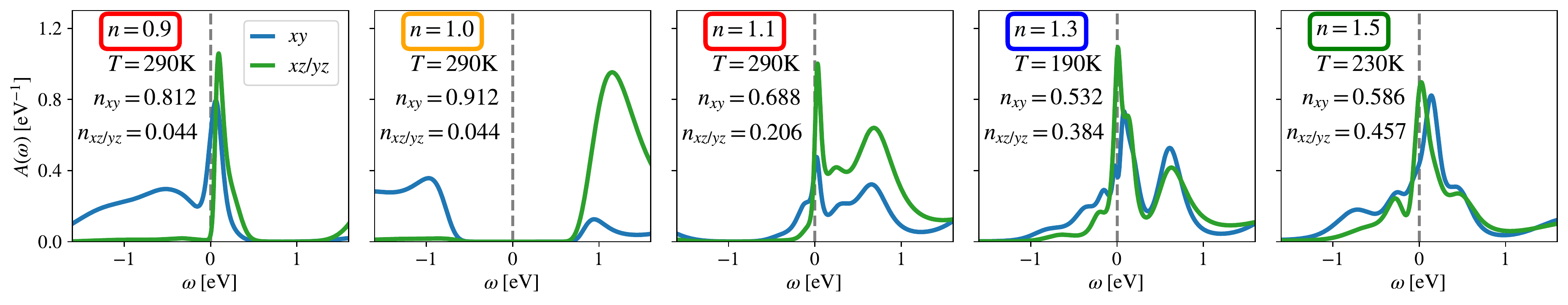}\\
  \caption{{\bf SrO-terminated monolayer -- DMFT spectral functions}  $A(\omega)$ for various fillings $n$ and temperatures $T$: in the Mott insulating state ($n=1.0$), in the vicinity of the AF ($n=0.9,1.1$), FM ($n=1.3$) and cOO ($n=1.5$) phases, resolved into orbital characters ($xy$  and $xz$/$yz$). 
  Colored boxes around fillings indicate the type of long-range orders realized at lower $T$, in correspondence to \fref{fig:phasediagram}.
  At nominal filling ($n=1$; orange) an orbitally polarized Mott insulator forms \cite{Pickem2021,PhysRevLett.114.246401}.
}
  \label{fig:spectral_sro}
\vspace{0.25cm}
    \includegraphics[width=1.0\textwidth]{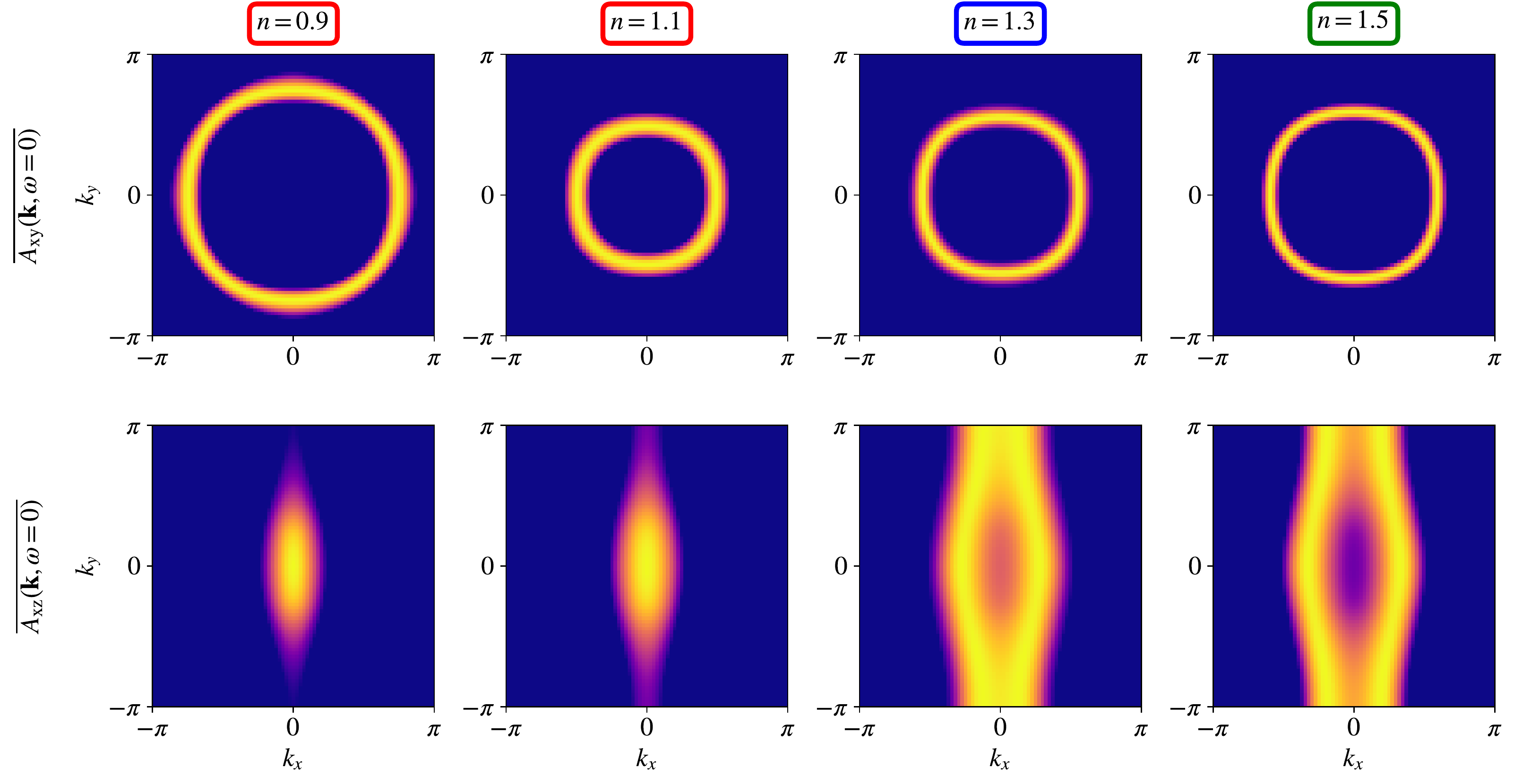}
  \caption{\textbf{SrO-terminated monolayer -- DMFT Fermi surface} for the points highlighted in Fig.~\ref{fig:phasediagram}a: $n=0.9$ ($T=290$K) and $n=1.1$ ($T=290$K) order AF  at low $T$ (red box indicating the color code of \fref{fig:phasediagram}), FM at $n=1.3$ ($T=190$K; blue box),  and cOO at  $n=1.5$ ($T=230$K; green box).}
  \label{fig:dmft_fermisurface_sro}
\end{figure*}

\subsection{SrO termination}
\label{sec:dmft_sro}

\fref{fig:dmft_fermisurface_sro} shows the DMFT Fermi surface 
for the SrO-terminated \svo\ monolayer at four different dopings (left to right). 
The upper panels display the contribution of the $xy$ orbital and the lower panels the $xz$ orbital (the  $yz$ orbital is equivalent to the latter if rotated by 90$^\circ$).

\begin{figure*}[tb]
  \includegraphics[width=1\textwidth]{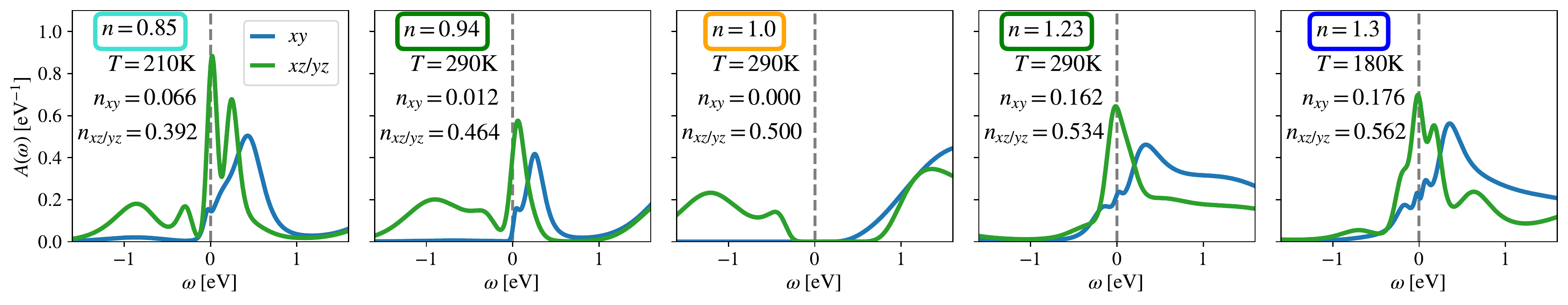}\\
  \caption{\textbf{VO$_2$-terminated monolayer -- DMFT spectral functions} $A(\omega)$ for various fillings $n$ and temperature $T$: $n=0.85$: sOO; $n=0.94$: cOO; $n=1.0$: Mott insulating cOO; $n=1.23$: cOO; $n=1.3$: iM. Otherwise identical to Fig.~\ref{fig:spectral_sro}.
  }
  \label{fig:spectral_vo2}
\vspace{0.25cm}
    \includegraphics[width=1.0\textwidth]{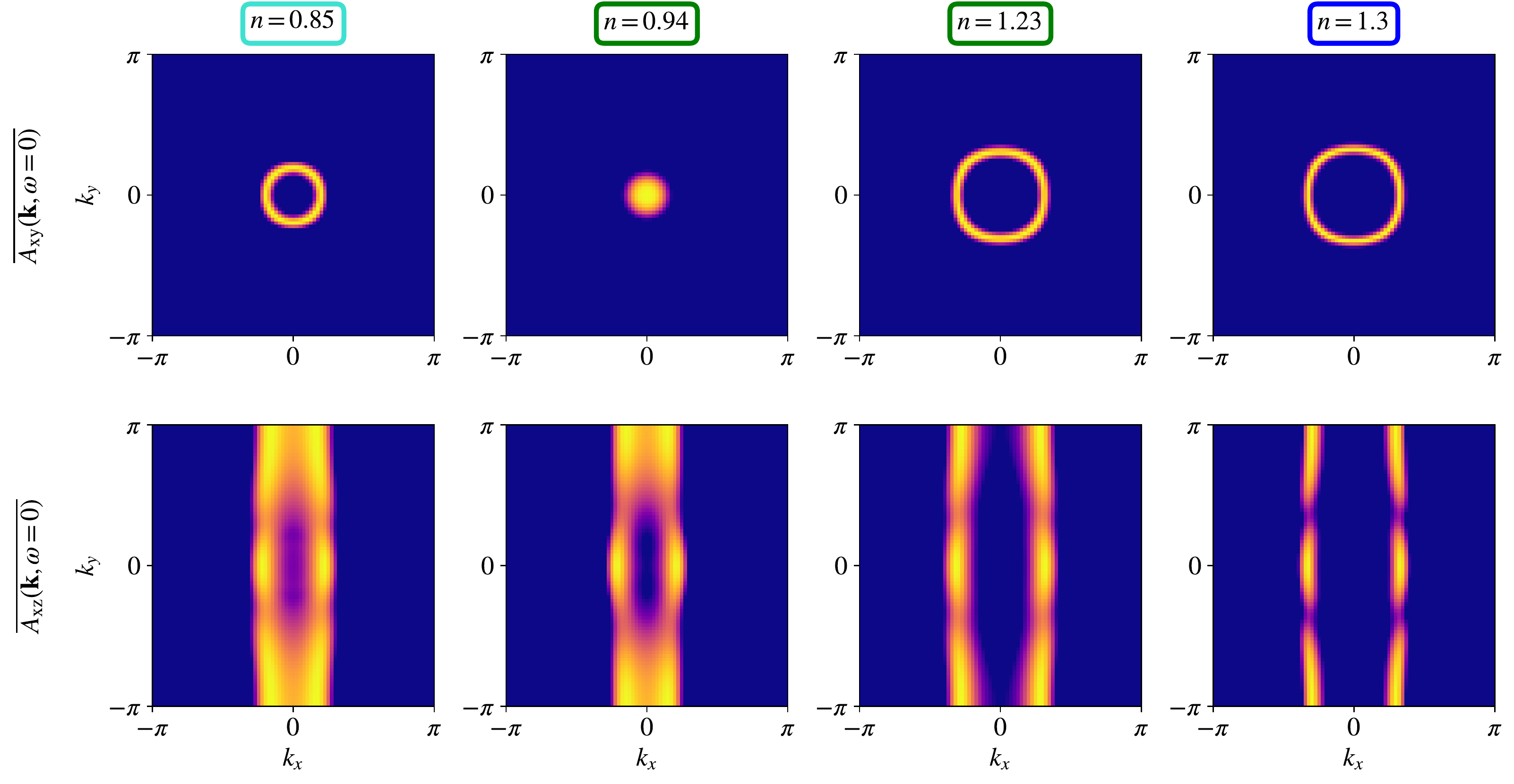}
  \caption{\textbf{VO$_2$-terminated monolayer -- DMFT Fermi surface}  for the points highlighted in Fig.~\ref{fig:phasediagram}b: $n=0.94$ ($T=290$K) and $n=1.23$ ($T=290$K) order cOO  at low $T$ (green box indicating the color code of \fref{fig:phasediagram}), $n=0.85$ sOO  ($T=210$K; turquoise box),  and $n=1.3$  iM ($T=190$K; blue box).}
  \label{fig:dmft_fermisurface_vo2}
\end{figure*}

We find the stoichiometric sample ($n=1$) to be an orbitally polarized insulator\cite{PhysRevLett.114.246401} with a gap of about 1$\,$eV, see \fref{fig:spectral_sro}.  That is, the in-plane ${xy}$ orbital is essentially half-filled, while the ${xz}$, ${yz}$ orbitals are almost completely depleted. Hence, the undoped SrO-terminated SrVO$_3$ monolayer is an effective one-orbital system.
The reduced orbital-degeneracy (with respect to the bulks threefold $t_{2g}$ orbitals) leads to a smaller critical interaction for the Mott state\cite{PhysRevB.54.R11026,pavarini:176403}. This turns the
undoped  SrVO$_3$ monolayer Mott insulating with strong antiferromagnetic (AF) fluctuations.

Doping with 10\% electrons or holes, we obtain a metal, see
the panels with $n=0.9$ and $n=1.1$, respectively, in \fref{fig:spectral_sro} and \fref{fig:dmft_fermisurface_sro}. The  ${xz}$ and ${yz}$ orbitals are now slightly filled, pushing
the AF phase transition to lower $T$, see \fref{fig:phasediagram}, while  strong AF spin fluctuations
persist. The  ${\mathbf k}$-integrated spectral function in 
\fref{fig:spectral_sro} further shows that the  ${xz}$ and ${yz}$ orbitals, while only slightly filled, already contribute a sizable amount to the quasi-particle peak at the Fermi level.

Similarities of this system to high-$T_c$ cuprates are uncanny:
While, here, the low-energy physics is dominated by
a half-filled ${xy}$ orbital instead of the ${x^2-y^2}$ orbital in cuprates, we find a ratio of nearest to next-neighbor in-plane hopping
$t\pr/t=+0.31$, ($t=-0.237\,$eV, $t'=-0.073\,$eV)
which is essentially the same as found for
YBa$_2$Cu$_3$O$_7$ and Bi$_2$Sr$_2$CaCu$_2$O$_8$ \cite{PhysRevLett.87.047003},
but $t'$ has the opposite sign. In a one-band picture, one can compensate for the opposite sign by making a particle-hole transformation and we obtain an electron-like Fermi surface instead of a hole like one in cuprates.
The decisive difference is however that, upon doping, the $xz$/$yz$ orbitals become partially
filled, leading to a different, multi-orbital kind of physics.

Indeed, at larger doping, $n=1.5$,    \fref{fig:phasediagram} indicates a checkerboard orbital-order (cOO) in
DMFT with a spatially alternating occupation of the $xz$ and $yz$ orbitals,
whereas the $xy$ orbital does not participate in the cOO.
Here, the $xy$ and $yz$ orbitals are already almost as much filled as the $xy$ orbital as is evident from \fref{fig:spectral_sro}  and also  from the Fermi surfaces in 
\fref{fig:dmft_fermisurface_sro}.
As the  $xz$ ($yz$) lobes point in the $x-$ ($y-$) and $z$-direction,
their Fermi surface in
\fref{fig:dmft_fermisurface_sro} is highly asymmetric, whereas their ${\mathbf k}$-integrated spectrum in \fref{fig:spectral_sro}  is similar to that of the $xy$ orbital.
In-between, around $n=1.3$, the $xz$ and $yz$ orbitals are still significantly less filled, however the spectral function at the Fermi level $A(\omega=0)$ is strongly enhanced, see \fref{fig:spectral_sro}. Ferromagnetic (FM) order therefore develops in \fref{fig:phasediagram} from the interplay of the Hund's coupling $J$ and the hopping $t$ \cite{Pickem2021}. 

\subsection{VO$_2$ termination}
\label{sec:dmft_vo2}

We now turn to the DMFT electronic structure of the VO$_2$-terminated surface. Again, we show Fermi surfaces 
(\fref{fig:dmft_fermisurface_vo2}) and ${\mathbf k}$-integrated spectra (\fref{fig:spectral_vo2}) for varying doping. For the VO$_2$- instead of the SrO-termination to the vacuum, the crystal-field splitting between the  $xz$/$yz$ and the $xy$ orbital flips its sign \cite{Pickem2021}. That is, the $xy$ orbital now lies above the
$xz$/$yz$ orbitals.
At $n=1$, the latter accommodate all of the charge and their
spectrum is split into upper and lower Hubbard bands, see  \fref{fig:spectral_vo2},
whereas the $xy$-orbital is unoccupied.
The two degenerate $xz$ and $yz$ orbitals are at or near quarter filling around $n=1$. This gives rise to checkerboard orbital fluctuations and, at low enough temperature, ordering (cOO) in DMFT, see \fref{fig:phasediagram}.
For slight hole doping and substantial electron doping, the cOO tendencies remain intact, but the \svo\ layer turns metallic.
Inverting the role of the   $xz$/$yz$ and the $xy$ orbitals compared to the SrO-termination, we now observe a small hole pocket for the $xy$ orbital in \fref{fig:dmft_fermisurface_vo2}, in agreement with their small filling in \fref{fig:spectral_vo2}.

Reducing the filling from $n=0.94$ to $n=0.85$, this $xy$ Fermi-surface pocket becomes slightly enhanced, albeit it remains small in \fref{fig:dmft_fermisurface_vo2}.
As for the fluctuations:
because of their reduced filling,
the $xz$/$yz$ orbitals are no longer quarter-filled. Thus cOO gives way, first, to stripe orbital order (sOO) at $n=0.85$ and, eventually, at lower fillings to FM,
similar as for 
the two-band Hubbard model \cite{Held1998}.

Further electron doping from $n=1.23$ to $n=1.3$ instead
changes the DMFT ordered state from cOO to incommensurate magnetism (iM) with a small ${\mathbf q}$-vector in \fref{fig:phasediagram}, see Ref.~\onlinecite{Pickem2021}. It has, however, little effect on the spectral function and the Fermi surface in  \fref{fig:spectral_vo2} and  \fref{fig:dmft_fermisurface_vo2}, respectively. The sharper Fermi surface for $n=1.3$ can be explained by the slight decrease in the temperature and the fact that {\em local} DMFT correlations get reduced the farther we are away from half-filling, see Fig.~\ref{fig:dga_fermisurface_vo2} below.

\begin{figure*}[htb]
    \includegraphics[width=0.9\textwidth]{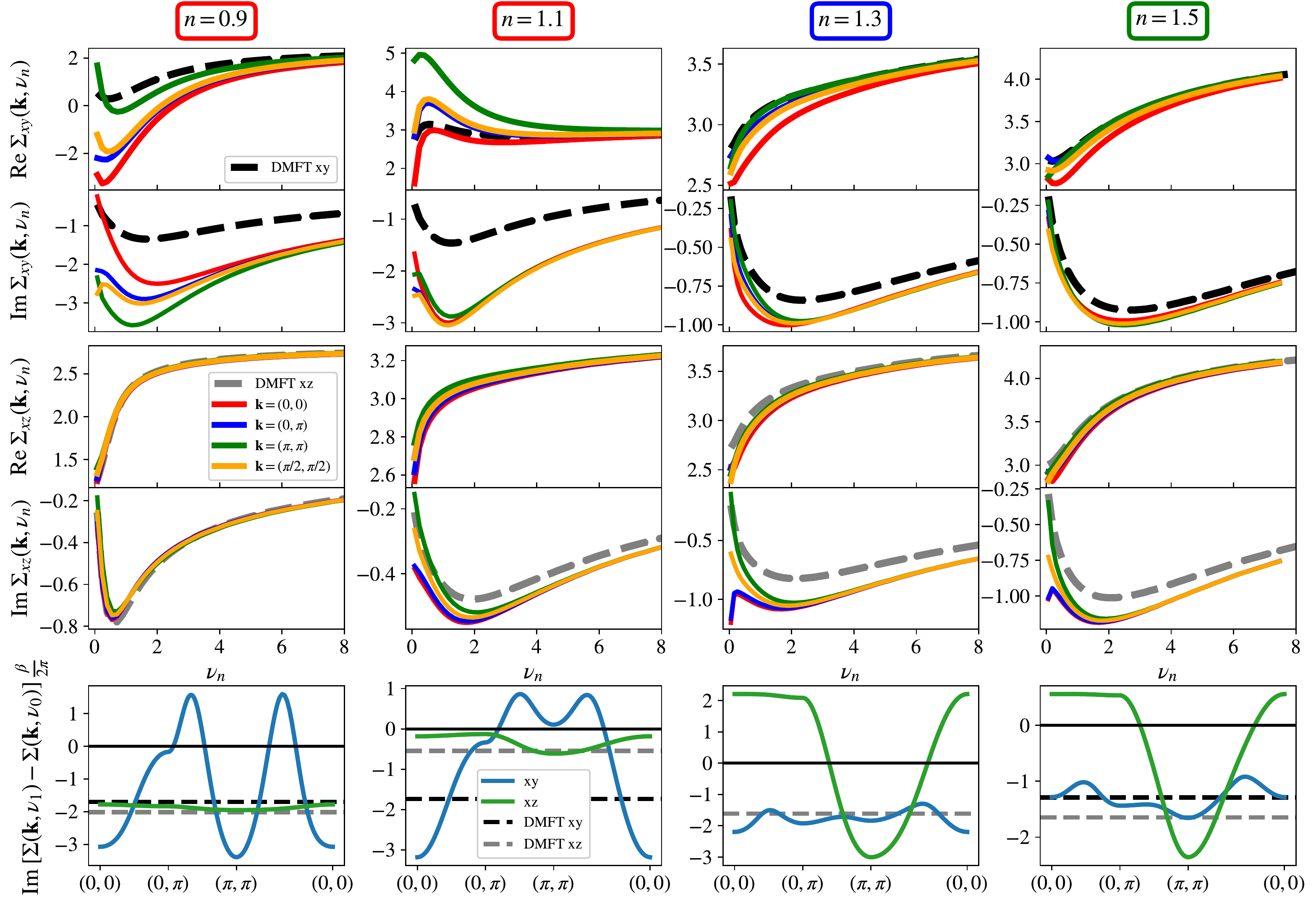}
  \caption{{\bf SrO-terminated monolayer -- Momentum differentiation of the D$\Gamma$A self-energy.} Top 4 rows:  real and imaginary parts for the $xy$ and $xz$ orbital at 4 different momenta, compared to DMFT, for the dopings and temperatures indicated in \fref{fig:phasediagram}a. Bottom row: Slope of the imaginary part of the  D$\Gamma$A self-energy for a path through the Brillouin zone. Negative values correspond to a Fermi-liquid like self-energy, positive values indicate the formation of a (pseudo)gap.}
  \label{fig:momentum_sro}
\end{figure*}
\begin{figure*}[htb]
    \includegraphics[width=0.9\textwidth]{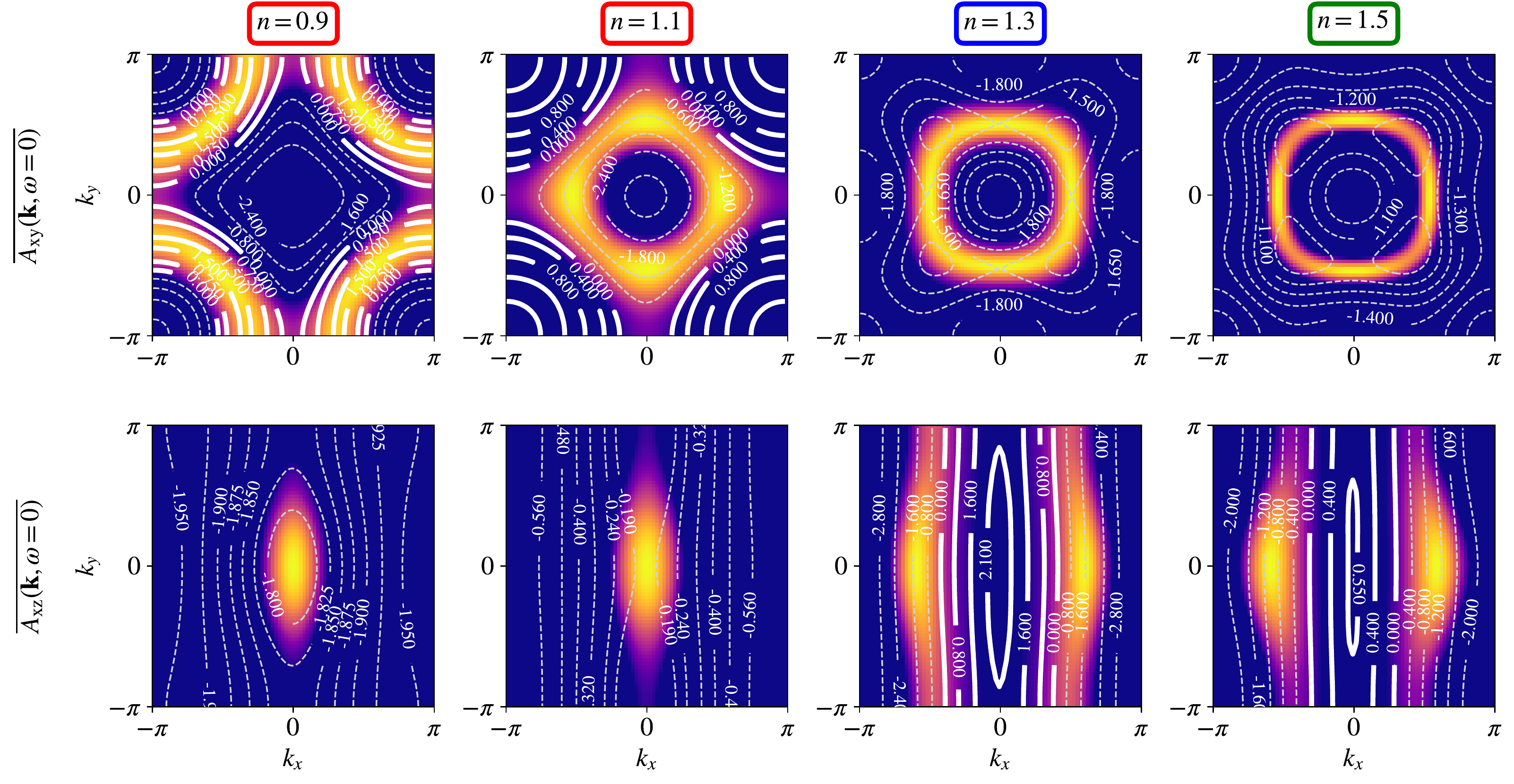}
  \caption{\textbf{SrO-terminated monolayer -- D$\Gamma$A Fermi surfaces} at the same dopings and $T$ as in Fig.~\ref{fig:dmft_fermisurface_sro}. The white contour lines represent isolines of the slope between the first two Matsubara frequencies of the imaginary part of $\Sigma$:  Solid, fat lines indicate a \emph{positive} slope, i.e., a kink in the self-energy, dashed, thin lines  a \emph{negative} value, suggestive of a Fermi-liquid-like state.
  }
  \label{fig:dga_fermisurface_sro_lifetime}
\end{figure*}

\section{D$\Gamma$A: momentum differentiation}
\label{Sec:ADGA}

On the dynamical mean-field level, many-body renormalizations are assumed to be isotropic (i.e., independent of momentum).
In 3D this is mostly a good approximation (see, however, Ref.~\onlinecite{jmt_dga3d}).
Yet, when the effective dimensionality is reduced, as in our ultrathin film, renormalizations become increasingly non-local \cite{Benjamin_2D3D}.
The major question we will answer here is:
{\it To what extent do the non-local critical fluctuations---in the vicinity of the associated
ordered states---lead to momentum-selective renormalizations?}
To elucidate this question, we use the AbinitioD$\Gamma$A\cite{Anna_ADGA, JPSJ_ADGA, CPC_ADGA} methodology
and scrutinize the electron self-energy $\Sigma(\vek{k},i\nu)$ in the vicinity of the DMFT ordering instabilities summarized above.

\subsection{SrO termination}

\fref{fig:momentum_sro} shows the real and imaginary part of the  AbinitioD$\Gamma$A self-energy in the vicinity of the DMFT phase transitions where non-local correlations become strong.
Shown are the two inequivalent orbitals, $xy$ (top) and $xz$ (middle panel) 
as a function of Matsubara frequency $\nu_n$. The $yz$ orbital is equivalent to the $xz$ orbital if the momenta are rotated by  90$^\circ$ rotated; the DMFT self-energy is shown for comparison.

In the vicinity of half-filling, $n=0.9$ and $n=1.1$, AF spin fluctuations   prevail with leading eigenvalue $\lambda_M(\pi,\pi) = 0.95$ and $0.79$, respectively, in the magnetic (M) channel at ${\mathbf q}=(\pi,\pi)$. Note, $\lambda=1$ indicates a divergence of the susceptibility, i.e., an ordering instability. These AF spin fluctuations are driven by the $xy$ orbital that is close to half filling, whereas the $xz$ and $yz$ orbitals rather act as passive bystanders \cite{Pickem2021}. Consequently, we see for $n=0.9$ and $n=1.1$ in \fref{fig:momentum_sro} a pronounced momentum differentiation only for the $xy$ orbital.

The Matsubara frequency self-energy has the advantage that it does not require the ill-conditioned analytic continuation. Nonetheless, we can gain valuable information: The momentum differentiation of the real part of the self-energy in \fref{fig:momentum_sro} between unoccupied [${\mathbf k}=(0,0)$, red]  and occupied states [${\mathbf k}=(\pi,\pi)$, green] signals that the quasi-particle poles at $\omega+\mu=\Re\Sigma+\epsilon_{\mathbf k}$ are pushed further away from the Fermi energy, causing an overall {\it enhancement} of the bandwidth.
The momentum differentiation between ${\mathbf k}=(0,\pi)$ (blue) and ${\mathbf k}=(\pi/2,\pi/2)$ (orange) that are closer to the Fermi level, indicates a deformation of the Fermi surface for $n=0.9$, but not for $n=1.1$ which has a similar self-energy for these two ${\mathbf k}$-points.
Indeed a deformation is observed in \fref{fig:dga_fermisurface_sro_lifetime}, where the electron-like DMFT Fermi surface (\fref{fig:dmft_fermisurface_sro}) turns into a hole-like one in D$\Gamma$A for $n=0.9$.
For $n=1.3$ with strong FM fluctuations ($\lambda_M(0,0) = 0.78$) and $n=1.5$ with strong cOO fluctuations in the density (D) channel ($\lambda_D(\pi,\pi) = 0.98 $), the momentum differentiation of $\Re\Sigma$ is less pronounced.

Let us now turn to $\Im\Sigma$ from which we can read off the scattering rate, as the $\nu_n\rightarrow 0$-extrapolated value. Further, from its slope the quasi-particle renormalization $Z_{\mathbf k}=[1-\partial \Im \Sigma({\mathbf k},i \nu)/\partial \nu|_{\nu\rightarrow 0}]^{-1}$
is accessible for a Fermi liquid phase.
A positive slope of $\Im \Sigma(i \nu\rightarrow 0)$ indicates the crossover to a diverging (Mott-like) self-energy, which splits the spectrum and leads to an insulating gap.

Clearly, for all four fillings shown in \fref{fig:momentum_sro}, there are momenta
for which the system exhibits non-Fermi liquid behavior,
identifiable by a kink and a downturn in $\Im\Sigma$ at low energies.
In case of AF fluctuations ($n=0.9$ and $n=1.1$) this downturn is in the $xy$ orbital, whereas it occurs in the $xz$ (and $yz$) orbital which dominates the  FM  ($n=1.3$) and cOO ($n=1.5$) fluctuations.
These kinks are salient indicators for the occurrence of a pseudogap state, and they get more pronounced when cooling the system toward the respective phase transition.

Interestingly, in the vicinity of the AF phase, the structure of the scattering rate is {\it opposite} 
to the cuprates: It is larger for the diagonal ($\pi$,$\pi$) direction than for the (0,$\pi$) direction. This momentum differentiation on the Fermi surface is, however, much less pronounced than the momentum dependence perpendicular to the Fermi surface, i.e., comparing occupied vs.\ unoccupied states.

This can be seen in \fref{fig:momentum_sro} (bottom), where we plot the slope between the first two positive Matsubara frequencies, i.e., $slope={(\Im\Sigma(i\nu_1) - \Im\Sigma(i\nu_0))}\beta/(2\pi)$,
 along the indicated $\vek{k}$-path.
Isolines of this slope are superimposed on the D$\Gamma$A Fermi surfaces in
\fref{fig:dga_fermisurface_sro_lifetime}, with the sign indicated by solid, fat (positive) and dashed, thin (negative) lines.
In the electron doped regime, the slope in $\Im\Sigma$ is always negative on the Fermi surface, i.e., Fermi liquid-like.
However, when moving away from the Fermi energy, we observe positive slopes, which corresponds to the kinks in
\fref{fig:momentum_sro};:
at  $n=1.1$  for the unoccupied $xy$ states above the Fermi level;  and at  $n=1.3$ and $n=1.5$ for the occupied $xz$ states.
In the hole doped regime, at $n=0.9$, we find $\Im\Sigma$ isoline patterns similar to $n=1.1$. However, owing to the larger $xy$-occupation in combination with the equally strong reconstruction through $\Re\Sigma$, negative slopes of $\Im\Sigma$ instead appear across the transformed $xy$ Fermi surface.
This insulating-like behavior is found only in the most relevant orbitals, i.e., the $xy$ orbital for the AF fluctuations around $n=1$, and the $xz/yz$ orbitals where FM and cOO long-range fluctuations are dominant.
The ancillary orbitals ($xz/yz$ for $n=1.1$ and $xy$ for $n=1.3,1.5$) on the other hand exhibit only a comparatively minor momentum differentiation (see \fref{fig:momentum_sro})---implying also a stark
orbital differentiation.

\begin{figure*}[tb]
    \includegraphics[width=0.9\textwidth]{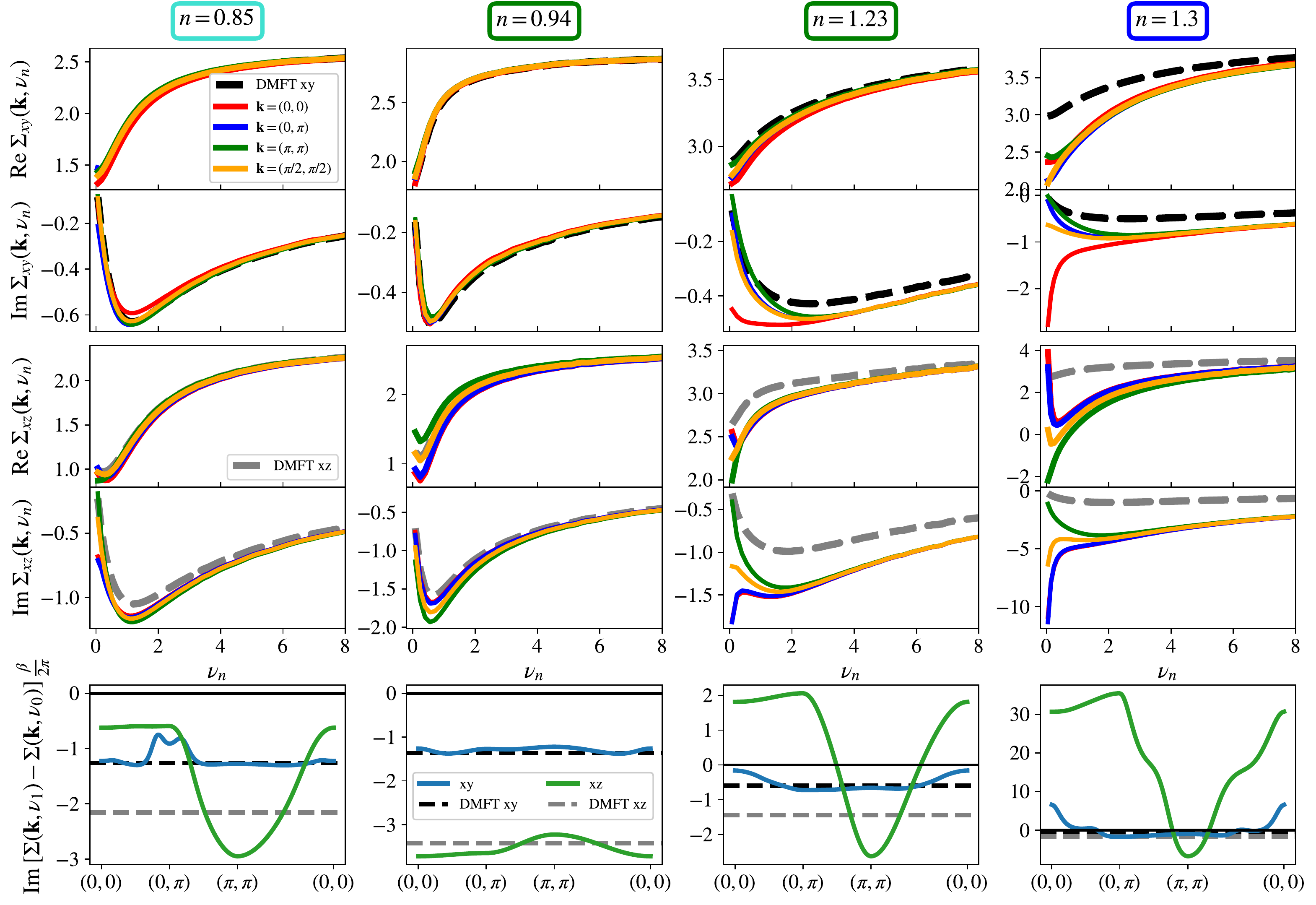}
    \caption{{\bf VO$_2$-terminated monolayer -- momentum differentiation of the D$\Gamma$A self-energy.} Top 4 rows: real and imaginary parts for the $xy$ and $xz$ orbital for 4 different momenta (colors) and, for comparison, the DMFT self-energy (dashed) at the four dopings and $T$ indicated by the ``+'' in \fref{fig:phasediagram}b. Bottom row: Slope of the imaginary part of the  D$\Gamma$A self-energy for a momentum path through the Brillouin zone.}
  \label{fig:momentum_vo2}
\end{figure*}
\begin{figure*}[tb]
    \includegraphics[width=0.9\textwidth]{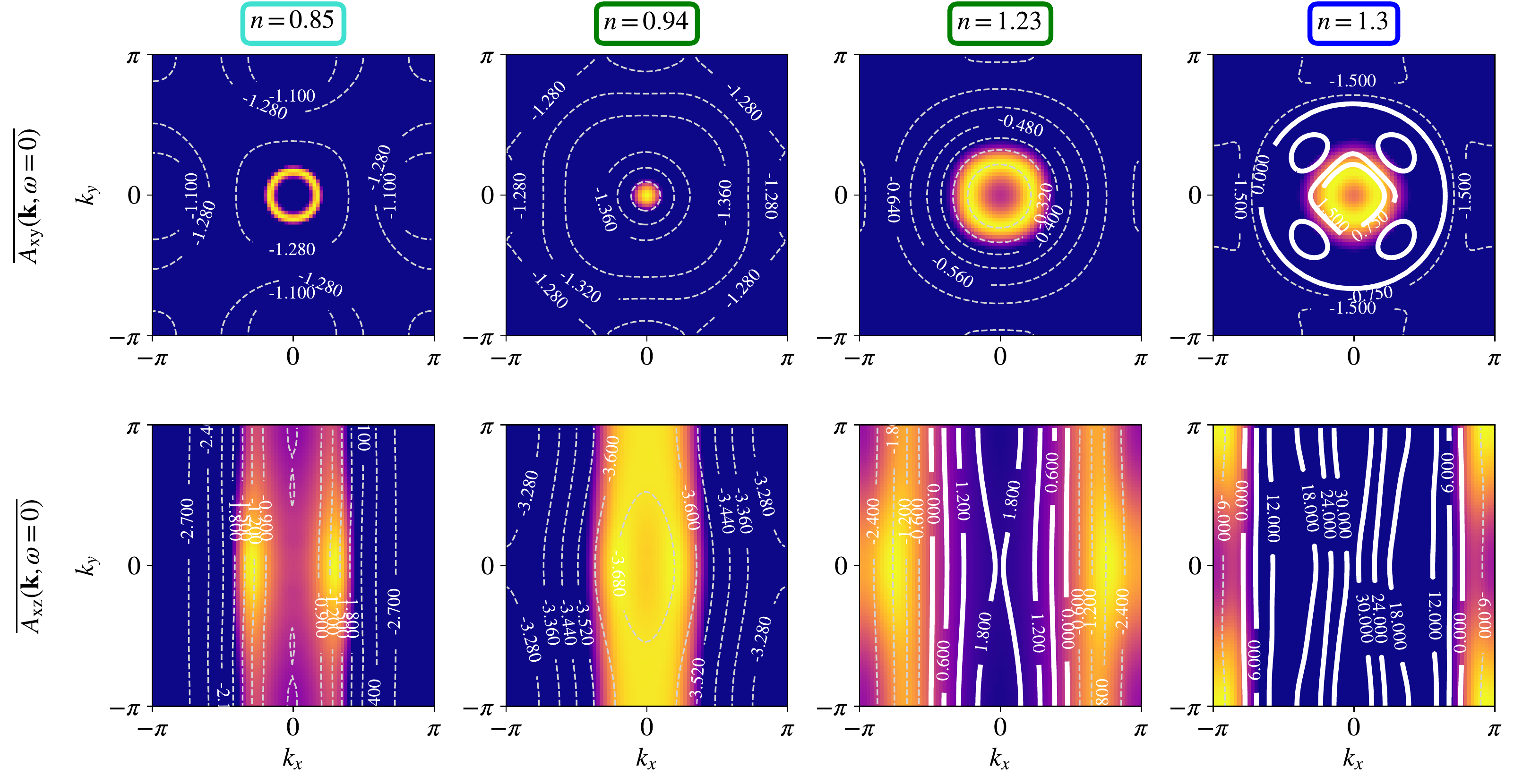}
  \caption{\textbf{VO$_2$-terminated monolayer -- D$\Gamma$A Fermi surfaces} at the same dopings and $T$ as in Fig.~\ref{fig:dmft_fermisurface_vo2}. Isolines are again the low-energy slope of $\Im\Sigma$, where the solid, fat lines represent a positive, and the dashed, thin lines represent a negative value.}
  \label{fig:dga_fermisurface_vo2}
\end{figure*}

\subsection{VO$_2$ termination}

The corresponding AbinitioD$\Gamma$A results for the  VO$_2$-terminated \svo\ monolayer for the self-energy and the Fermi surface are presented in \fref{fig:momentum_vo2}
and \fref{fig:dga_fermisurface_vo2}, respectively.
For cOO fluctuations at $n=1.23$  ($\lambda_D(\pi,\pi) = 0.97$), the momentum differentiation of the self-energy and Fermi surface are qualitatively similar to the cOO results at $n=1.5$ for the SrO-terminated layer. But for the cOO at $n=0.94$ and sOO at $n=0.85$,  we only find a minor momentum differentiation of the  self-energy, see \fref{fig:momentum_vo2}. Correspondingly, the Fermi surface in 
\fref{fig:dga_fermisurface_vo2} is similar to that of DMFT in \fref{fig:dmft_fermisurface_vo2}, and there are no positive non-Fermi-liquid like slopes (solid lines in \fref{fig:dga_fermisurface_vo2}). This is surprising since the leading eigenvalue $\lambda_D(0,\pi) = \lambda_D(\pi,0) = 0.985$ at $n=0.85$
and  $\lambda_D(\pi,\pi) = 0.91$ at $n=0.94$ is similarly close to 1 as for $n=1.23$ or the SrO-termination, indicating that strong orbital ordering fluctuations are present.

On the contrary, at $n=1.3$, above iM order,  we observe the by far strongest momentum differentiation in
\fref{fig:momentum_vo2},
even though  $\lambda_M(\delta,\delta) = 0.97$ with $\delta \approx \pm \pi/4$ is again comparable to the strength of other fluctuations.
A clear pole develops in the vicinity of the Fermi level not only for  the $xz$ and $yz$ orbitals, that drive the iM ordering, but also for the $xy$ orbital.
This pole is so large that the spectrum splits into two parts, akin to the splitting into upper and lower Hubbard band; and it pushes the Fermi surface 
to ${\mathbf k}=(0,\pm\pi)$.
However, the divergence occurs only for a region of the Brillouin zone that does not account for the Fermi surface of 
the respective orbital character.


\section{Discussion and Perspective}
\label{Sec:discussion}

Recapitulating, 
we have studied a \svo\ monolayer on a SrTiO$_3$ substrate with two different surface terminations, SrO and VO$_2$, to vacuum within AbinitioD$\Gamma$A. Depending on the termination
and filling, there are strong non-local fluctuations of various kinds:
antiferromagnetic, ferromagnetic, incommensurate magnetic, striped or checkerboard orbital. These non-local fluctuations will suppress the mean-field DMFT ordering but also have pronounced effects on the self-energy---the focus of the present paper. They can deform the Fermi surface, as observed for antiferromagnetic ordering with $n=0.9$ for SrO-terminated \svo, and quite generally can lead to a strong enhancement of $\Im\Sigma$.
Strong non-local fluctuations can even cause the development of a pole in the self-energy, signaling the splitting of the spectrum into two parts---here not because of Mott-Hubbard physics but because of large non-local fluctuations.
The latter is particularly strong for the incommensurate ferromagnetic phase of the VO$_2$-terminated \svo\ monolayer at $n=1.3$ filling. First indications, i.e.,  downturns of the self-energy at the lowest Matsubara frequency are however ubiquitous for various dopings and both terminations.
While such non-local physics have been investigated quite intensively for antiferromagnetic fluctuations in the Hubbard model in the context of the cuprates, to the best of our knowledge it has not been analyzed before for orbital fluctuations.

The undoped ($n=1$) SrVO$_3$ monolayer is Mott insulating and, for SrO-termination, appears to be akin to the cuprates with the $xy$ orbital playing the role of the high-$T_c$'s $x^2-y^2$ orbital.
However, electron-doping reveals the multi-orbital physics of the SrVO$_3$ system: The $xy$ orbital is depopulated when adding electrons to the system, and all three orbitals, $xy$, $xz$ and $yz$, participate in developing a quasi-particle resonance at the Fermi level.

For the cuprates, AF fluctuations lead to pseudogap physics  with a momentum differentiation distinguishing between a Fermi liquid-like self-energy in the nodal direction on the Fermi surface, and a kink in the self-energy signaling the opening of a gap in the anti-nodal direction. Here, we also observe the joint presence of these two behaviors in the self-energy. However, the momentum differentiation is not realized {\it on} the Fermi surface but {\it perpendicular} to it: For the SrO-termination in the electron-doped regime, AF fluctuations lead to a Fermi liquid-like behavior for momenta on the occupied side of the Fermi surface
("$k<k_F$") and a kink-like insulating behavior in the imaginary part of the self-energy on the unoccupied side ("$k>k_F$").  In case of FM and OO fluctuations, the momentum differentiation between occupied and unoccupied momenta is reversed. 

For the VO$_2$ termination,  iM fluctuations at $n=1.3$ lead to massive non-local correlations and a pole in the self-energy. In contrast to all other cases not only the $xz$/$yz$ orbitals---driving the iM fluctuations---are affected but also the ancillary $xy$ orbital. Below half filling, on the other hand,
cOO and sOO fluctuations only result in a minor momentum differentiation of the self-energy for the VO$_2$-terminated monolayer.

The imaginary part of the self-energy corresponds to the lifetimes and the broadening of the spectral function. Our results hence show that the lifetimes of an added hole or electron are extremely different. The hole-lifetime can be measured by angular resolved photoemission spectroscopy (ARPES); the electron lifetime by inverse photoemission spectroscopy, by ARPES at elevated temperatures, or in non-equilibrium situations (e.g., pump-probe measurements) in which states above the Fermi level become populated.

The differentiation between states above and below the Fermi surface that we observe is quite extreme. Technologically this might be exploited for thermoelectrics which rely on a  strong electron-hole asymmetry \cite{Mahan07231996,karsten_hvar,zlatic,NGCS}.
Particularly beneficial are sharp peaks in the spectral function on only one side of the Fermi level \cite{Mahan07231996}, as found for the SrO-terminated monolayer, see \fref{fig:dmft_fermisurface_sro}, within DMFT. There, {\it local} electronic correlations can enhance thermoelectricity through energy-dependent renormalizations that are different for electrons ($\omega<0$) and holes ($\omega<0)$\cite{10.1007/978-90-481-2892-1_7,karsten_hvar}.
Our finding of a momentum-selectivity in the scattering rate may provide an {\em additional} route:
A particle-hole asymmetry that is driven (or enhanced) by {\it non-local} renormalizations.
Indeed, looking again at the SrO-terminated monolayer, 
OO and FM fluctuations at $n=1.5$ and $n=1.3$, respectively, drive 
a dispersive scattering rate\cite{PhysRevB.88.245203} that is larger for occupied momenta ("$k<k_F$") then for empty states ("$k>k_F$"):
Specifically, the downward kinks in the $xz$-component of $\Im\Sigma$, see \fref{fig:momentum_sro}, occur for $\mathbf{k}=(0,0)$ and $\mathbf{k}=(0,\pi)$
which are inside the (DMFT) $xz$ Fermi surface, see \fref{fig:dmft_fermisurface_sro}.
For $\mathbf{k}=(\pi,\pi)$ and $\mathbf{k}=(\pi/2,\pi/2)$, which are outside the Fermi surface, the $xz$ scattering rate instead decreases when approaching zero frequency.
This electron-hole asymmetry of the scattering time will make the already electron-like DMFT thermopower even more negative, thus increasing its magnitude.

\section{Conclusions}
\label{Sec:Conclusion}
Based on simulations of oxide ultrathin films,
we demonstrated that, in (quasi-) two-dimensional systems, strong long-range fluctuations are quite generically reflected in a momentum differentiation of the self-energy, irrespective of the dominant fluctuation channel (spin/charge/orbital).
Further, we demonstrated that this momentum differentiation has a much richer structure than the current focus on cuprates and pseudogap physics suggests: Strong variations in renormalizations may not only occur on the Fermi surface but also perpendicular to it.
Our results call for a (re)examining---with beyond-DMFT methods---of correlated electron systems that host strong non-local fluctuations: layered materials as well as ultra-thin oxide films and heterostructures.

\begin{acknowledgments}
  We thank M.\ Fuchs, A.\ Galler, J. Kaufmann,  G.\ Sangiovanni,  P.\ Thunstr\"om and Z.\ Zhong for fruitful discussions.
 The authors acknowledge support from the Austrian Science Fund (FWF) through grants P 30819,  P 32044,   and P 30213. Calculations were performed on the Vienna Scientific Cluster (VSC).
 \end{acknowledgments}


\begin{thebibliography}{103}%
\makeatletter
\providecommand \@ifxundefined [1]{%
 \@ifx{#1\undefined}
}%
\providecommand \@ifnum [1]{%
 \ifnum #1\expandafter \@firstoftwo
 \else \expandafter \@secondoftwo
 \fi
}%
\providecommand \@ifx [1]{%
 \ifx #1\expandafter \@firstoftwo
 \else \expandafter \@secondoftwo
 \fi
}%
\providecommand \natexlab [1]{#1}%
\providecommand \enquote  [1]{``#1''}%
\providecommand \bibnamefont  [1]{#1}%
\providecommand \bibfnamefont [1]{#1}%
\providecommand \citenamefont [1]{#1}%
\providecommand \href@noop [0]{\@secondoftwo}%
\providecommand \href [0]{\begingroup \@sanitize@url \@href}%
\providecommand \@href[1]{\@@startlink{#1}\@@href}%
\providecommand \@@href[1]{\endgroup#1\@@endlink}%
\providecommand \@sanitize@url [0]{\catcode `\\12\catcode `\$12\catcode
  `\&12\catcode `\#12\catcode `\^12\catcode `\_12\catcode `\%12\relax}%
\providecommand \@@startlink[1]{}%
\providecommand \@@endlink[0]{}%
\providecommand \url  [0]{\begingroup\@sanitize@url \@url }%
\providecommand \@url [1]{\endgroup\@href {#1}{\urlprefix }}%
\providecommand \urlprefix  [0]{URL }%
\providecommand \Eprint [0]{\href }%
\providecommand \doibase [0]{http://dx.doi.org/}%
\providecommand \selectlanguage [0]{\@gobble}%
\providecommand \bibinfo  [0]{\@secondoftwo}%
\providecommand \bibfield  [0]{\@secondoftwo}%
\providecommand \translation [1]{[#1]}%
\providecommand \BibitemOpen [0]{}%
\providecommand \bibitemStop [0]{}%
\providecommand \bibitemNoStop [0]{.\EOS\space}%
\providecommand \EOS [0]{\spacefactor3000\relax}%
\providecommand \BibitemShut  [1]{\csname bibitem#1\endcsname}%
\let\auto@bib@innerbib\@empty
\bibitem [{\citenamefont {Vilk}\ and\ \citenamefont
  {Tremblay}(1997)}]{Vilk1997}%
  \BibitemOpen
  \bibfield  {author} {\bibinfo {author} {\bibfnamefont {Y.~M.}\ \bibnamefont
  {Vilk}}\ and\ \bibinfo {author} {\bibfnamefont {A.-M.~S.}\ \bibnamefont
  {Tremblay}},\ }\href {\doibase 10.1051/jp1:1997135} {\bibfield  {journal}
  {\bibinfo  {journal} {J. Phys. I France}\ }\textbf {\bibinfo {volume} {7}},\
  \bibinfo {pages} {1309} (\bibinfo {year} {1997})}\BibitemShut {NoStop}%
\bibitem [{\citenamefont {Rohringer}\ \emph {et~al.}(2011)\citenamefont
  {Rohringer}, \citenamefont {Toschi}, \citenamefont {Katanin},\ and\
  \citenamefont {Held}}]{Rohringer2011}%
  \BibitemOpen
  \bibfield  {author} {\bibinfo {author} {\bibfnamefont {G.}~\bibnamefont
  {Rohringer}}, \bibinfo {author} {\bibfnamefont {A.}~\bibnamefont {Toschi}},
  \bibinfo {author} {\bibfnamefont {A.}~\bibnamefont {Katanin}}, \ and\
  \bibinfo {author} {\bibfnamefont {K.}~\bibnamefont {Held}},\ }\href {\doibase
  10.1103/PhysRevLett.107.256402} {\bibfield  {journal} {\bibinfo  {journal}
  {Phys. Rev. Lett.}\ }\textbf {\bibinfo {volume} {107}},\ \bibinfo {pages}
  {256402} (\bibinfo {year} {2011})}\BibitemShut {NoStop}%
\bibitem [{\citenamefont {Rohringer}\ and\ \citenamefont
  {Toschi}(2016)}]{Rohringer2016}%
  \BibitemOpen
  \bibfield  {author} {\bibinfo {author} {\bibfnamefont {G.}~\bibnamefont
  {Rohringer}}\ and\ \bibinfo {author} {\bibfnamefont {A.}~\bibnamefont
  {Toschi}},\ }\href {\doibase 10.1103/PhysRevB.94.125144} {\bibfield
  {journal} {\bibinfo  {journal} {Phys. Rev. B}\ }\textbf {\bibinfo {volume}
  {94}},\ \bibinfo {pages} {125144} (\bibinfo {year} {2016})}\BibitemShut
  {NoStop}%
\bibitem [{\citenamefont {Norman}\ \emph {et~al.}(1998)\citenamefont {Norman},
  \citenamefont {Ding}, \citenamefont {Randeria}, \citenamefont {Campuzano},
  \citenamefont {Yokoya}, \citenamefont {Takeuchi}, \citenamefont {Takahashi},
  \citenamefont {Mochiku}, \citenamefont {Kadowaki}, \citenamefont
  {Guptasarma},\ and\ \citenamefont {Hinks}}]{Norman1998}%
  \BibitemOpen
  \bibfield  {author} {\bibinfo {author} {\bibfnamefont {M.~R.}\ \bibnamefont
  {Norman}}, \bibinfo {author} {\bibfnamefont {H.}~\bibnamefont {Ding}},
  \bibinfo {author} {\bibfnamefont {M.}~\bibnamefont {Randeria}}, \bibinfo
  {author} {\bibfnamefont {J.~C.}\ \bibnamefont {Campuzano}}, \bibinfo {author}
  {\bibfnamefont {T.}~\bibnamefont {Yokoya}}, \bibinfo {author} {\bibfnamefont
  {T.}~\bibnamefont {Takeuchi}}, \bibinfo {author} {\bibfnamefont
  {T.}~\bibnamefont {Takahashi}}, \bibinfo {author} {\bibfnamefont
  {T.}~\bibnamefont {Mochiku}}, \bibinfo {author} {\bibfnamefont
  {K.}~\bibnamefont {Kadowaki}}, \bibinfo {author} {\bibfnamefont
  {P.}~\bibnamefont {Guptasarma}}, \ and\ \bibinfo {author} {\bibfnamefont
  {D.~G.}\ \bibnamefont {Hinks}},\ }\href {\doibase 10.1038/32366} {\bibfield
  {journal} {\bibinfo  {journal} {Nature}\ }\textbf {\bibinfo {volume} {392}},\
  \bibinfo {pages} {157} (\bibinfo {year} {1998})}\BibitemShut {NoStop}%
\bibitem [{\citenamefont {Timusk}\ and\ \citenamefont
  {Statt}(1999)}]{Timusk_1999}%
  \BibitemOpen
  \bibfield  {author} {\bibinfo {author} {\bibfnamefont {T.}~\bibnamefont
  {Timusk}}\ and\ \bibinfo {author} {\bibfnamefont {B.}~\bibnamefont {Statt}},\
  }\href {\doibase 10.1088/0034-4885/62/1/002} {\bibfield  {journal} {\bibinfo
  {journal} {Reports on Progress in Physics}\ }\textbf {\bibinfo {volume}
  {62}},\ \bibinfo {pages} {61} (\bibinfo {year} {1999})}\BibitemShut {NoStop}%
\bibitem [{\citenamefont {Norman}\ \emph {et~al.}(2005)\citenamefont {Norman},
  \citenamefont {Pines},\ and\ \citenamefont {Kallin}}]{Norman2005}%
  \BibitemOpen
  \bibfield  {author} {\bibinfo {author} {\bibfnamefont {M.~R.}\ \bibnamefont
  {Norman}}, \bibinfo {author} {\bibfnamefont {D.}~\bibnamefont {Pines}}, \
  and\ \bibinfo {author} {\bibfnamefont {C.}~\bibnamefont {Kallin}},\ }\href
  {\doibase 10.1080/00018730500459906} {\bibfield  {journal} {\bibinfo
  {journal} {Advances in Physics}\ }\textbf {\bibinfo {volume} {54}},\ \bibinfo
  {pages} {715} (\bibinfo {year} {2005})}\BibitemShut {NoStop}%
\bibitem [{\citenamefont {Keimer}\ \emph {et~al.}(2015)\citenamefont {Keimer},
  \citenamefont {Kivelson}, \citenamefont {Norman}, \citenamefont {Uchida},\
  and\ \citenamefont {Zaanen}}]{Keimer2015}%
  \BibitemOpen
  \bibfield  {author} {\bibinfo {author} {\bibfnamefont {B.}~\bibnamefont
  {Keimer}}, \bibinfo {author} {\bibfnamefont {S.~A.}\ \bibnamefont
  {Kivelson}}, \bibinfo {author} {\bibfnamefont {M.~R.}\ \bibnamefont
  {Norman}}, \bibinfo {author} {\bibfnamefont {S.}~\bibnamefont {Uchida}}, \
  and\ \bibinfo {author} {\bibfnamefont {J.}~\bibnamefont {Zaanen}},\ }\href
  {\doibase 10.1038/nature14165} {\bibfield  {journal} {\bibinfo  {journal}
  {Nature}\ }\textbf {\bibinfo {volume} {518}},\ \bibinfo {pages} {179}
  (\bibinfo {year} {2015})}\BibitemShut {NoStop}%
\bibitem [{\citenamefont {Kampf}\ and\ \citenamefont
  {Schrieffer}(1990{\natexlab{a}})}]{PhysRevB.42.7967}%
  \BibitemOpen
  \bibfield  {author} {\bibinfo {author} {\bibfnamefont {A.~P.}\ \bibnamefont
  {Kampf}}\ and\ \bibinfo {author} {\bibfnamefont {J.~R.}\ \bibnamefont
  {Schrieffer}},\ }\href {\doibase 10.1103/PhysRevB.42.7967} {\bibfield
  {journal} {\bibinfo  {journal} {Phys. Rev. B}\ }\textbf {\bibinfo {volume}
  {42}},\ \bibinfo {pages} {7967} (\bibinfo {year}
  {1990}{\natexlab{a}})}\BibitemShut {NoStop}%
\bibitem [{\citenamefont {Vilk}\ and\ \citenamefont
  {Tremblay}(1996)}]{Vilk_1996}%
  \BibitemOpen
  \bibfield  {author} {\bibinfo {author} {\bibfnamefont {Y.~M.}\ \bibnamefont
  {Vilk}}\ and\ \bibinfo {author} {\bibfnamefont {A.-M.~S.}\ \bibnamefont
  {Tremblay}},\ }\href {\doibase 10.1209/epl/i1996-00315-2} {\bibfield
  {journal} {\bibinfo  {journal} {Europhysics Letters ({EPL})}\ }\textbf
  {\bibinfo {volume} {33}},\ \bibinfo {pages} {159} (\bibinfo {year}
  {1996})}\BibitemShut {NoStop}%
\bibitem [{\citenamefont {Rost}\ \emph {et~al.}(2012)\citenamefont {Rost},
  \citenamefont {Gorelik}, \citenamefont {Assaad},\ and\ \citenamefont
  {Bl\"umer}}]{Rost2012}%
  \BibitemOpen
  \bibfield  {author} {\bibinfo {author} {\bibfnamefont {D.}~\bibnamefont
  {Rost}}, \bibinfo {author} {\bibfnamefont {E.~V.}\ \bibnamefont {Gorelik}},
  \bibinfo {author} {\bibfnamefont {F.}~\bibnamefont {Assaad}}, \ and\ \bibinfo
  {author} {\bibfnamefont {N.}~\bibnamefont {Bl\"umer}},\ }\href {\doibase
  10.1103/PhysRevB.86.155109} {\bibfield  {journal} {\bibinfo  {journal} {Phys.
  Rev. B}\ }\textbf {\bibinfo {volume} {86}},\ \bibinfo {pages} {155109}
  (\bibinfo {year} {2012})}\BibitemShut {NoStop}%
\bibitem [{\citenamefont {Kyung}\ \emph {et~al.}(2004)\citenamefont {Kyung},
  \citenamefont {Hankevych}, \citenamefont {Dar\'e},\ and\ \citenamefont
  {Tremblay}}]{PhysRevLett.93.147004}%
  \BibitemOpen
  \bibfield  {author} {\bibinfo {author} {\bibfnamefont {B.}~\bibnamefont
  {Kyung}}, \bibinfo {author} {\bibfnamefont {V.}~\bibnamefont {Hankevych}},
  \bibinfo {author} {\bibfnamefont {A.-M.}\ \bibnamefont {Dar\'e}}, \ and\
  \bibinfo {author} {\bibfnamefont {A.-M.~S.}\ \bibnamefont {Tremblay}},\
  }\href {\doibase 10.1103/PhysRevLett.93.147004} {\bibfield  {journal}
  {\bibinfo  {journal} {Phys. Rev. Lett.}\ }\textbf {\bibinfo {volume} {93}},\
  \bibinfo {pages} {147004} (\bibinfo {year} {2004})}\BibitemShut {NoStop}%
\bibitem [{\citenamefont {Gull}\ \emph {et~al.}(2013)\citenamefont {Gull},
  \citenamefont {Parcollet},\ and\ \citenamefont {Millis}}]{Gull2013}%
  \BibitemOpen
  \bibfield  {author} {\bibinfo {author} {\bibfnamefont {E.}~\bibnamefont
  {Gull}}, \bibinfo {author} {\bibfnamefont {O.}~\bibnamefont {Parcollet}}, \
  and\ \bibinfo {author} {\bibfnamefont {A.~J.}\ \bibnamefont {Millis}},\
  }\href {\doibase 10.1103/PhysRevLett.110.216405} {\bibfield  {journal}
  {\bibinfo  {journal} {Phys. Rev. Lett.}\ }\textbf {\bibinfo {volume} {110}},\
  \bibinfo {pages} {216405} (\bibinfo {year} {2013})}\BibitemShut {NoStop}%
\bibitem [{\citenamefont {Cyr-Choini\`ere}\ \emph {et~al.}(2018)\citenamefont
  {Cyr-Choini\`ere}, \citenamefont {Daou}, \citenamefont {Lalibert\'e},
  \citenamefont {Collignon}, \citenamefont {Badoux}, \citenamefont {LeBoeuf},
  \citenamefont {Chang}, \citenamefont {Ramshaw}, \citenamefont {Bonn},
  \citenamefont {Hardy}, \citenamefont {Liang}, \citenamefont {Yan},
  \citenamefont {Cheng}, \citenamefont {Zhou}, \citenamefont {Goodenough},
  \citenamefont {Pyon}, \citenamefont {Takayama}, \citenamefont {Takagi},
  \citenamefont {Doiron-Leyraud},\ and\ \citenamefont
  {Taillefer}}]{Cyr-Choiniere18}%
  \BibitemOpen
  \bibfield  {author} {\bibinfo {author} {\bibfnamefont {O.}~\bibnamefont
  {Cyr-Choini\`ere}}, \bibinfo {author} {\bibfnamefont {R.}~\bibnamefont
  {Daou}}, \bibinfo {author} {\bibfnamefont {F.}~\bibnamefont {Lalibert\'e}},
  \bibinfo {author} {\bibfnamefont {C.}~\bibnamefont {Collignon}}, \bibinfo
  {author} {\bibfnamefont {S.}~\bibnamefont {Badoux}}, \bibinfo {author}
  {\bibfnamefont {D.}~\bibnamefont {LeBoeuf}}, \bibinfo {author} {\bibfnamefont
  {J.}~\bibnamefont {Chang}}, \bibinfo {author} {\bibfnamefont {B.~J.}\
  \bibnamefont {Ramshaw}}, \bibinfo {author} {\bibfnamefont {D.~A.}\
  \bibnamefont {Bonn}}, \bibinfo {author} {\bibfnamefont {W.~N.}\ \bibnamefont
  {Hardy}}, \bibinfo {author} {\bibfnamefont {R.}~\bibnamefont {Liang}},
  \bibinfo {author} {\bibfnamefont {J.-Q.}\ \bibnamefont {Yan}}, \bibinfo
  {author} {\bibfnamefont {J.-G.}\ \bibnamefont {Cheng}}, \bibinfo {author}
  {\bibfnamefont {J.-S.}\ \bibnamefont {Zhou}}, \bibinfo {author}
  {\bibfnamefont {J.~B.}\ \bibnamefont {Goodenough}}, \bibinfo {author}
  {\bibfnamefont {S.}~\bibnamefont {Pyon}}, \bibinfo {author} {\bibfnamefont
  {T.}~\bibnamefont {Takayama}}, \bibinfo {author} {\bibfnamefont
  {H.}~\bibnamefont {Takagi}}, \bibinfo {author} {\bibfnamefont
  {N.}~\bibnamefont {Doiron-Leyraud}}, \ and\ \bibinfo {author} {\bibfnamefont
  {L.}~\bibnamefont {Taillefer}},\ }\href {\doibase 10.1103/PhysRevB.97.064502}
  {\bibfield  {journal} {\bibinfo  {journal} {Phys. Rev. B}\ }\textbf {\bibinfo
  {volume} {97}},\ \bibinfo {pages} {064502} (\bibinfo {year}
  {2018})}\BibitemShut {NoStop}%
\bibitem [{\citenamefont {Kampf}\ and\ \citenamefont
  {Schrieffer}(1990{\natexlab{b}})}]{Kampf90}%
  \BibitemOpen
  \bibfield  {author} {\bibinfo {author} {\bibfnamefont {A.}~\bibnamefont
  {Kampf}}\ and\ \bibinfo {author} {\bibfnamefont {J.~R.}\ \bibnamefont
  {Schrieffer}},\ }\href {\doibase 10.1103/PhysRevB.41.6399} {\bibfield
  {journal} {\bibinfo  {journal} {Phys. Rev. B}\ }\textbf {\bibinfo {volume}
  {41}},\ \bibinfo {pages} {6399} (\bibinfo {year}
  {1990}{\natexlab{b}})}\BibitemShut {NoStop}%
\bibitem [{\citenamefont {Monthoux}\ and\ \citenamefont
  {Pines}(1993)}]{Pines93}%
  \BibitemOpen
  \bibfield  {author} {\bibinfo {author} {\bibfnamefont {P.}~\bibnamefont
  {Monthoux}}\ and\ \bibinfo {author} {\bibfnamefont {D.}~\bibnamefont
  {Pines}},\ }\href {\doibase 10.1103/PhysRevB.47.6069} {\bibfield  {journal}
  {\bibinfo  {journal} {Phys. Rev. B}\ }\textbf {\bibinfo {volume} {47}},\
  \bibinfo {pages} {6069} (\bibinfo {year} {1993})}\BibitemShut {NoStop}%
\bibitem [{\citenamefont {Abanov}\ \emph {et~al.}(2003)\citenamefont {Abanov},
  \citenamefont {Chubukov},\ and\ \citenamefont {Schmalian}}]{Abanov03}%
  \BibitemOpen
  \bibfield  {author} {\bibinfo {author} {\bibfnamefont {A.}~\bibnamefont
  {Abanov}}, \bibinfo {author} {\bibfnamefont {A.~V.}\ \bibnamefont
  {Chubukov}}, \ and\ \bibinfo {author} {\bibfnamefont {J.}~\bibnamefont
  {Schmalian}},\ }\href {\doibase 10.1080/0001873021000057123} {\bibfield
  {journal} {\bibinfo  {journal} {Advances in Physics}\ }\textbf {\bibinfo
  {volume} {52}},\ \bibinfo {pages} {119} (\bibinfo {year} {2003})}\BibitemShut
  {NoStop}%
\bibitem [{\citenamefont {Vilk}(1997)}]{Vilk97-2}%
  \BibitemOpen
  \bibfield  {author} {\bibinfo {author} {\bibfnamefont {Y.~M.}\ \bibnamefont
  {Vilk}},\ }\href {\doibase 10.1103/PhysRevB.55.3870} {\bibfield  {journal}
  {\bibinfo  {journal} {Phys. Rev. B}\ }\textbf {\bibinfo {volume} {55}},\
  \bibinfo {pages} {3870} (\bibinfo {year} {1997})}\BibitemShut {NoStop}%
\bibitem [{\citenamefont {Wu}\ \emph {et~al.}(2020{\natexlab{a}})\citenamefont
  {Wu}, \citenamefont {Scheurer}, \citenamefont {Ferrero},\ and\ \citenamefont
  {Georges}}]{Wu20}%
  \BibitemOpen
  \bibfield  {author} {\bibinfo {author} {\bibfnamefont {W.}~\bibnamefont
  {Wu}}, \bibinfo {author} {\bibfnamefont {M.~S.}\ \bibnamefont {Scheurer}},
  \bibinfo {author} {\bibfnamefont {M.}~\bibnamefont {Ferrero}}, \ and\
  \bibinfo {author} {\bibfnamefont {A.}~\bibnamefont {Georges}},\ }\href
  {\doibase 10.1103/PhysRevResearch.2.033067} {\bibfield  {journal} {\bibinfo
  {journal} {Phys. Rev. Research}\ }\textbf {\bibinfo {volume} {2}},\ \bibinfo
  {pages} {033067} (\bibinfo {year} {2020}{\natexlab{a}})}\BibitemShut
  {NoStop}%
\bibitem [{\citenamefont {Gonz\'alez}\ \emph {et~al.}(2000)\citenamefont
  {Gonz\'alez}, \citenamefont {Guinea},\ and\ \citenamefont
  {Vozmediano}}]{Gonzalez00}%
  \BibitemOpen
  \bibfield  {author} {\bibinfo {author} {\bibfnamefont {J.}~\bibnamefont
  {Gonz\'alez}}, \bibinfo {author} {\bibfnamefont {F.}~\bibnamefont {Guinea}},
  \ and\ \bibinfo {author} {\bibfnamefont {M.~A.~H.}\ \bibnamefont
  {Vozmediano}},\ }\href {\doibase 10.1103/PhysRevLett.84.4930} {\bibfield
  {journal} {\bibinfo  {journal} {Phys. Rev. Lett.}\ }\textbf {\bibinfo
  {volume} {84}},\ \bibinfo {pages} {4930} (\bibinfo {year}
  {2000})}\BibitemShut {NoStop}%
\bibitem [{\citenamefont {Halboth}\ and\ \citenamefont
  {Metzner}(2000)}]{Halboth20b}%
  \BibitemOpen
  \bibfield  {author} {\bibinfo {author} {\bibfnamefont {C.~J.}\ \bibnamefont
  {Halboth}}\ and\ \bibinfo {author} {\bibfnamefont {W.}~\bibnamefont
  {Metzner}},\ }\href {\doibase 10.1103/PhysRevLett.85.5162} {\bibfield
  {journal} {\bibinfo  {journal} {Phys. Rev. Lett.}\ }\textbf {\bibinfo
  {volume} {85}},\ \bibinfo {pages} {5162} (\bibinfo {year}
  {2000})}\BibitemShut {NoStop}%
\bibitem [{\citenamefont {Honerkamp}\ and\ \citenamefont
  {Salmhofer}(2001)}]{Honerkamp01}%
  \BibitemOpen
  \bibfield  {author} {\bibinfo {author} {\bibfnamefont {C.}~\bibnamefont
  {Honerkamp}}\ and\ \bibinfo {author} {\bibfnamefont {M.}~\bibnamefont
  {Salmhofer}},\ }\href {\doibase 10.1103/PhysRevLett.87.187004} {\bibfield
  {journal} {\bibinfo  {journal} {Phys. Rev. Lett.}\ }\textbf {\bibinfo
  {volume} {87}},\ \bibinfo {pages} {187004} (\bibinfo {year}
  {2001})}\BibitemShut {NoStop}%
\bibitem [{\citenamefont {Wu}\ \emph {et~al.}(2020{\natexlab{b}})\citenamefont
  {Wu}, \citenamefont {Scheurer}, \citenamefont {Ferrero},\ and\ \citenamefont
  {Georges}}]{PhysRevResearch.2.033067}%
  \BibitemOpen
  \bibfield  {author} {\bibinfo {author} {\bibfnamefont {W.}~\bibnamefont
  {Wu}}, \bibinfo {author} {\bibfnamefont {M.~S.}\ \bibnamefont {Scheurer}},
  \bibinfo {author} {\bibfnamefont {M.}~\bibnamefont {Ferrero}}, \ and\
  \bibinfo {author} {\bibfnamefont {A.}~\bibnamefont {Georges}},\ }\href
  {\doibase 10.1103/PhysRevResearch.2.033067} {\bibfield  {journal} {\bibinfo
  {journal} {Phys. Rev. Research}\ }\textbf {\bibinfo {volume} {2}},\ \bibinfo
  {pages} {033067} (\bibinfo {year} {2020}{\natexlab{b}})}\BibitemShut
  {NoStop}%
\bibitem [{\citenamefont {{Krien}}\ \emph {et~al.}(2021)\citenamefont
  {{Krien}}, \citenamefont {{Worm}}, \citenamefont {{Chalupa}}, \citenamefont
  {{Toschi}},\ and\ \citenamefont {{Held}}}]{Krien2021}%
  \BibitemOpen
  \bibfield  {author} {\bibinfo {author} {\bibfnamefont {F.}~\bibnamefont
  {{Krien}}}, \bibinfo {author} {\bibfnamefont {P.}~\bibnamefont {{Worm}}},
  \bibinfo {author} {\bibfnamefont {P.}~\bibnamefont {{Chalupa}}}, \bibinfo
  {author} {\bibfnamefont {A.}~\bibnamefont {{Toschi}}}, \ and\ \bibinfo
  {author} {\bibfnamefont {K.}~\bibnamefont {{Held}}},\ }\href
  {https://ui.adsabs.harvard.edu/abs/2021arXiv210706529K} {\bibfield  {journal}
  {\bibinfo  {journal} {arXiv:2107.06529}\ } (\bibinfo {year}
  {2021})}\BibitemShut {NoStop}%
\bibitem [{\citenamefont {Fay}\ \emph {et~al.}(1988)\citenamefont {Fay},
  \citenamefont {Loesener},\ and\ \citenamefont {Appel}}]{PhysRevB.37.3299}%
  \BibitemOpen
  \bibfield  {author} {\bibinfo {author} {\bibfnamefont {D.}~\bibnamefont
  {Fay}}, \bibinfo {author} {\bibfnamefont {O.}~\bibnamefont {Loesener}}, \
  and\ \bibinfo {author} {\bibfnamefont {J.}~\bibnamefont {Appel}},\ }\href
  {\doibase 10.1103/PhysRevB.37.3299} {\bibfield  {journal} {\bibinfo
  {journal} {Phys. Rev. B}\ }\textbf {\bibinfo {volume} {37}},\ \bibinfo
  {pages} {3299} (\bibinfo {year} {1988})}\BibitemShut {NoStop}%
\bibitem [{\citenamefont {Monthoux}(2003)}]{PhysRevB.68.064408}%
  \BibitemOpen
  \bibfield  {author} {\bibinfo {author} {\bibfnamefont {P.}~\bibnamefont
  {Monthoux}},\ }\href {\doibase 10.1103/PhysRevB.68.064408} {\bibfield
  {journal} {\bibinfo  {journal} {Phys. Rev. B}\ }\textbf {\bibinfo {volume}
  {68}},\ \bibinfo {pages} {064408} (\bibinfo {year} {2003})}\BibitemShut
  {NoStop}%
\bibitem [{\citenamefont {Hankevych}\ \emph {et~al.}(2003)\citenamefont
  {Hankevych}, \citenamefont {Kyung},\ and\ \citenamefont
  {Tremblay}}]{PhysRevB.68.214405}%
  \BibitemOpen
  \bibfield  {author} {\bibinfo {author} {\bibfnamefont {V.}~\bibnamefont
  {Hankevych}}, \bibinfo {author} {\bibfnamefont {B.}~\bibnamefont {Kyung}}, \
  and\ \bibinfo {author} {\bibfnamefont {A.-M.~S.}\ \bibnamefont {Tremblay}},\
  }\href {\doibase 10.1103/PhysRevB.68.214405} {\bibfield  {journal} {\bibinfo
  {journal} {Phys. Rev. B}\ }\textbf {\bibinfo {volume} {68}},\ \bibinfo
  {pages} {214405} (\bibinfo {year} {2003})}\BibitemShut {NoStop}%
\bibitem [{\citenamefont {Katanin}\ \emph {et~al.}(2005)\citenamefont
  {Katanin}, \citenamefont {Kampf},\ and\ \citenamefont
  {Irkhin}}]{PhysRevB.71.085105}%
  \BibitemOpen
  \bibfield  {author} {\bibinfo {author} {\bibfnamefont {A.~A.}\ \bibnamefont
  {Katanin}}, \bibinfo {author} {\bibfnamefont {A.~P.}\ \bibnamefont {Kampf}},
  \ and\ \bibinfo {author} {\bibfnamefont {V.~Y.}\ \bibnamefont {Irkhin}},\
  }\href {\doibase 10.1103/PhysRevB.71.085105} {\bibfield  {journal} {\bibinfo
  {journal} {Phys. Rev. B}\ }\textbf {\bibinfo {volume} {71}},\ \bibinfo
  {pages} {085105} (\bibinfo {year} {2005})}\BibitemShut {NoStop}%
\bibitem [{\citenamefont {Katanin}(2005)}]{PhysRevB.72.035111}%
  \BibitemOpen
  \bibfield  {author} {\bibinfo {author} {\bibfnamefont {A.~A.}\ \bibnamefont
  {Katanin}},\ }\href {\doibase 10.1103/PhysRevB.72.035111} {\bibfield
  {journal} {\bibinfo  {journal} {Phys. Rev. B}\ }\textbf {\bibinfo {volume}
  {72}},\ \bibinfo {pages} {035111} (\bibinfo {year} {2005})}\BibitemShut
  {NoStop}%
\bibitem [{\citenamefont {Pickem}\ \emph {et~al.}(2021)\citenamefont {Pickem},
  \citenamefont {Kaufmann}, \citenamefont {Held},\ and\ \citenamefont
  {Tomczak}}]{Pickem2021}%
  \BibitemOpen
  \bibfield  {author} {\bibinfo {author} {\bibfnamefont {M.}~\bibnamefont
  {Pickem}}, \bibinfo {author} {\bibfnamefont {J.}~\bibnamefont {Kaufmann}},
  \bibinfo {author} {\bibfnamefont {K.}~\bibnamefont {Held}}, \ and\ \bibinfo
  {author} {\bibfnamefont {J.~M.}\ \bibnamefont {Tomczak}},\ }\href {\doibase
  10.1103/PhysRevB.104.024307} {\bibfield  {journal} {\bibinfo  {journal}
  {Phys. Rev. B}\ }\textbf {\bibinfo {volume} {104}},\ \bibinfo {pages}
  {024307} (\bibinfo {year} {2021})}\BibitemShut {NoStop}%
\bibitem [{\citenamefont {Klebel-Knobloch}\ \emph {et~al.}(2021)\citenamefont
  {Klebel-Knobloch}, \citenamefont {Sch\"afer}, \citenamefont {Toschi},\ and\
  \citenamefont {Tomczak}}]{Benjamin_2D3D}%
  \BibitemOpen
  \bibfield  {author} {\bibinfo {author} {\bibfnamefont {B.}~\bibnamefont
  {Klebel-Knobloch}}, \bibinfo {author} {\bibfnamefont {T.}~\bibnamefont
  {Sch\"afer}}, \bibinfo {author} {\bibfnamefont {A.}~\bibnamefont {Toschi}}, \
  and\ \bibinfo {author} {\bibfnamefont {J.~M.}\ \bibnamefont {Tomczak}},\
  }\href {\doibase 10.1103/PhysRevB.103.045121} {\bibfield  {journal} {\bibinfo
   {journal} {Phys. Rev. B}\ }\textbf {\bibinfo {volume} {103}},\ \bibinfo
  {pages} {045121} (\bibinfo {year} {2021})}\BibitemShut {NoStop}%
\bibitem [{\citenamefont {Xu}\ \emph {et~al.}(2011)\citenamefont {Xu},
  \citenamefont {Richard}, \citenamefont {Nakayama}, \citenamefont {Kawahara},
  \citenamefont {Sekiba}, \citenamefont {Qian}, \citenamefont {Neupane},
  \citenamefont {Souma}, \citenamefont {Sato}, \citenamefont {Takahashi},
  \citenamefont {Luo}, \citenamefont {Wen}, \citenamefont {Chen}, \citenamefont
  {Wang}, \citenamefont {Wang}, \citenamefont {Fang}, \citenamefont {Dai},\
  and\ \citenamefont {Ding}}]{Xu2011}%
  \BibitemOpen
  \bibfield  {author} {\bibinfo {author} {\bibfnamefont {Y.-M.}\ \bibnamefont
  {Xu}}, \bibinfo {author} {\bibfnamefont {P.}~\bibnamefont {Richard}},
  \bibinfo {author} {\bibfnamefont {K.}~\bibnamefont {Nakayama}}, \bibinfo
  {author} {\bibfnamefont {T.}~\bibnamefont {Kawahara}}, \bibinfo {author}
  {\bibfnamefont {Y.}~\bibnamefont {Sekiba}}, \bibinfo {author} {\bibfnamefont
  {T.}~\bibnamefont {Qian}}, \bibinfo {author} {\bibfnamefont {M.}~\bibnamefont
  {Neupane}}, \bibinfo {author} {\bibfnamefont {S.}~\bibnamefont {Souma}},
  \bibinfo {author} {\bibfnamefont {T.}~\bibnamefont {Sato}}, \bibinfo {author}
  {\bibfnamefont {T.}~\bibnamefont {Takahashi}}, \bibinfo {author}
  {\bibfnamefont {H.-Q.}\ \bibnamefont {Luo}}, \bibinfo {author} {\bibfnamefont
  {H.-H.}\ \bibnamefont {Wen}}, \bibinfo {author} {\bibfnamefont {G.-F.}\
  \bibnamefont {Chen}}, \bibinfo {author} {\bibfnamefont {N.-L.}\ \bibnamefont
  {Wang}}, \bibinfo {author} {\bibfnamefont {Z.}~\bibnamefont {Wang}}, \bibinfo
  {author} {\bibfnamefont {Z.}~\bibnamefont {Fang}}, \bibinfo {author}
  {\bibfnamefont {X.}~\bibnamefont {Dai}}, \ and\ \bibinfo {author}
  {\bibfnamefont {H.}~\bibnamefont {Ding}},\ }\href {\doibase
  10.1038/ncomms1394} {\bibfield  {journal} {\bibinfo  {journal} {Nature
  Communications}\ }\textbf {\bibinfo {volume} {2}},\ \bibinfo {pages} {392}
  (\bibinfo {year} {2011})}\BibitemShut {NoStop}%
\bibitem [{\citenamefont {Moon}\ \emph {et~al.}(2012)\citenamefont {Moon},
  \citenamefont {Schafgans}, \citenamefont {Kasahara}, \citenamefont
  {Shibauchi}, \citenamefont {Terashima}, \citenamefont {Matsuda},
  \citenamefont {Tanatar}, \citenamefont {Prozorov}, \citenamefont {Thaler},
  \citenamefont {Canfield}, \citenamefont {Sefat}, \citenamefont {Mandrus},\
  and\ \citenamefont {Basov}}]{PhysRevLett.109.027006}%
  \BibitemOpen
  \bibfield  {author} {\bibinfo {author} {\bibfnamefont {S.~J.}\ \bibnamefont
  {Moon}}, \bibinfo {author} {\bibfnamefont {A.~A.}\ \bibnamefont {Schafgans}},
  \bibinfo {author} {\bibfnamefont {S.}~\bibnamefont {Kasahara}}, \bibinfo
  {author} {\bibfnamefont {T.}~\bibnamefont {Shibauchi}}, \bibinfo {author}
  {\bibfnamefont {T.}~\bibnamefont {Terashima}}, \bibinfo {author}
  {\bibfnamefont {Y.}~\bibnamefont {Matsuda}}, \bibinfo {author} {\bibfnamefont
  {M.~A.}\ \bibnamefont {Tanatar}}, \bibinfo {author} {\bibfnamefont
  {R.}~\bibnamefont {Prozorov}}, \bibinfo {author} {\bibfnamefont
  {A.}~\bibnamefont {Thaler}}, \bibinfo {author} {\bibfnamefont {P.~C.}\
  \bibnamefont {Canfield}}, \bibinfo {author} {\bibfnamefont {A.~S.}\
  \bibnamefont {Sefat}}, \bibinfo {author} {\bibfnamefont {D.}~\bibnamefont
  {Mandrus}}, \ and\ \bibinfo {author} {\bibfnamefont {D.~N.}\ \bibnamefont
  {Basov}},\ }\href {\doibase 10.1103/PhysRevLett.109.027006} {\bibfield
  {journal} {\bibinfo  {journal} {Phys. Rev. Lett.}\ }\textbf {\bibinfo
  {volume} {109}},\ \bibinfo {pages} {027006} (\bibinfo {year}
  {2012})}\BibitemShut {NoStop}%
\bibitem [{\citenamefont {Zhou}\ \emph {et~al.}(2012)\citenamefont {Zhou},
  \citenamefont {Cai}, \citenamefont {Wang}, \citenamefont {Ruan},
  \citenamefont {Ye}, \citenamefont {Chen}, \citenamefont {You}, \citenamefont
  {Weng},\ and\ \citenamefont {Wang}}]{PhysRevLett.109.037002}%
  \BibitemOpen
  \bibfield  {author} {\bibinfo {author} {\bibfnamefont {X.}~\bibnamefont
  {Zhou}}, \bibinfo {author} {\bibfnamefont {P.}~\bibnamefont {Cai}}, \bibinfo
  {author} {\bibfnamefont {A.}~\bibnamefont {Wang}}, \bibinfo {author}
  {\bibfnamefont {W.}~\bibnamefont {Ruan}}, \bibinfo {author} {\bibfnamefont
  {C.}~\bibnamefont {Ye}}, \bibinfo {author} {\bibfnamefont {X.}~\bibnamefont
  {Chen}}, \bibinfo {author} {\bibfnamefont {Y.}~\bibnamefont {You}}, \bibinfo
  {author} {\bibfnamefont {Z.-Y.}\ \bibnamefont {Weng}}, \ and\ \bibinfo
  {author} {\bibfnamefont {Y.}~\bibnamefont {Wang}},\ }\href {\doibase
  10.1103/PhysRevLett.109.037002} {\bibfield  {journal} {\bibinfo  {journal}
  {Phys. Rev. Lett.}\ }\textbf {\bibinfo {volume} {109}},\ \bibinfo {pages}
  {037002} (\bibinfo {year} {2012})}\BibitemShut {NoStop}%
\bibitem [{\citenamefont {Shimojima}\ \emph {et~al.}(2014)\citenamefont
  {Shimojima}, \citenamefont {Sonobe}, \citenamefont {Malaeb}, \citenamefont
  {Shinada}, \citenamefont {Chainani}, \citenamefont {Shin}, \citenamefont
  {Yoshida}, \citenamefont {Ideta}, \citenamefont {Fujimori}, \citenamefont
  {Kumigashira}, \citenamefont {Ono}, \citenamefont {Nakashima}, \citenamefont
  {Anzai}, \citenamefont {Arita}, \citenamefont {Ino}, \citenamefont
  {Namatame}, \citenamefont {Taniguchi}, \citenamefont {Nakajima},
  \citenamefont {Uchida}, \citenamefont {Tomioka}, \citenamefont {Ito},
  \citenamefont {Kihou}, \citenamefont {Lee}, \citenamefont {Iyo},
  \citenamefont {Eisaki}, \citenamefont {Ohgushi}, \citenamefont {Kasahara},
  \citenamefont {Terashima}, \citenamefont {Ikeda}, \citenamefont {Shibauchi},
  \citenamefont {Matsuda},\ and\ \citenamefont
  {Ishizaka}}]{PhysRevB.89.045101}%
  \BibitemOpen
  \bibfield  {author} {\bibinfo {author} {\bibfnamefont {T.}~\bibnamefont
  {Shimojima}}, \bibinfo {author} {\bibfnamefont {T.}~\bibnamefont {Sonobe}},
  \bibinfo {author} {\bibfnamefont {W.}~\bibnamefont {Malaeb}}, \bibinfo
  {author} {\bibfnamefont {K.}~\bibnamefont {Shinada}}, \bibinfo {author}
  {\bibfnamefont {A.}~\bibnamefont {Chainani}}, \bibinfo {author}
  {\bibfnamefont {S.}~\bibnamefont {Shin}}, \bibinfo {author} {\bibfnamefont
  {T.}~\bibnamefont {Yoshida}}, \bibinfo {author} {\bibfnamefont
  {S.}~\bibnamefont {Ideta}}, \bibinfo {author} {\bibfnamefont
  {A.}~\bibnamefont {Fujimori}}, \bibinfo {author} {\bibfnamefont
  {H.}~\bibnamefont {Kumigashira}}, \bibinfo {author} {\bibfnamefont
  {K.}~\bibnamefont {Ono}}, \bibinfo {author} {\bibfnamefont {Y.}~\bibnamefont
  {Nakashima}}, \bibinfo {author} {\bibfnamefont {H.}~\bibnamefont {Anzai}},
  \bibinfo {author} {\bibfnamefont {M.}~\bibnamefont {Arita}}, \bibinfo
  {author} {\bibfnamefont {A.}~\bibnamefont {Ino}}, \bibinfo {author}
  {\bibfnamefont {H.}~\bibnamefont {Namatame}}, \bibinfo {author}
  {\bibfnamefont {M.}~\bibnamefont {Taniguchi}}, \bibinfo {author}
  {\bibfnamefont {M.}~\bibnamefont {Nakajima}}, \bibinfo {author}
  {\bibfnamefont {S.}~\bibnamefont {Uchida}}, \bibinfo {author} {\bibfnamefont
  {Y.}~\bibnamefont {Tomioka}}, \bibinfo {author} {\bibfnamefont
  {T.}~\bibnamefont {Ito}}, \bibinfo {author} {\bibfnamefont {K.}~\bibnamefont
  {Kihou}}, \bibinfo {author} {\bibfnamefont {C.~H.}\ \bibnamefont {Lee}},
  \bibinfo {author} {\bibfnamefont {A.}~\bibnamefont {Iyo}}, \bibinfo {author}
  {\bibfnamefont {H.}~\bibnamefont {Eisaki}}, \bibinfo {author} {\bibfnamefont
  {K.}~\bibnamefont {Ohgushi}}, \bibinfo {author} {\bibfnamefont
  {S.}~\bibnamefont {Kasahara}}, \bibinfo {author} {\bibfnamefont
  {T.}~\bibnamefont {Terashima}}, \bibinfo {author} {\bibfnamefont
  {H.}~\bibnamefont {Ikeda}}, \bibinfo {author} {\bibfnamefont
  {T.}~\bibnamefont {Shibauchi}}, \bibinfo {author} {\bibfnamefont
  {Y.}~\bibnamefont {Matsuda}}, \ and\ \bibinfo {author} {\bibfnamefont
  {K.}~\bibnamefont {Ishizaka}},\ }\href {\doibase 10.1103/PhysRevB.89.045101}
  {\bibfield  {journal} {\bibinfo  {journal} {Phys. Rev. B}\ }\textbf {\bibinfo
  {volume} {89}},\ \bibinfo {pages} {045101} (\bibinfo {year}
  {2014})}\BibitemShut {NoStop}%
\bibitem [{\citenamefont {Lin}\ \emph {et~al.}(2013)\citenamefont {Lin},
  \citenamefont {Texier}, \citenamefont {Taleb-Ibrahimi}, \citenamefont
  {Le~F\`evre}, \citenamefont {Bertran}, \citenamefont {Giannini},
  \citenamefont {Grioni},\ and\ \citenamefont
  {Brouet}}]{PhysRevLett.111.217002}%
  \BibitemOpen
  \bibfield  {author} {\bibinfo {author} {\bibfnamefont {P.-H.}\ \bibnamefont
  {Lin}}, \bibinfo {author} {\bibfnamefont {Y.}~\bibnamefont {Texier}},
  \bibinfo {author} {\bibfnamefont {A.}~\bibnamefont {Taleb-Ibrahimi}},
  \bibinfo {author} {\bibfnamefont {P.}~\bibnamefont {Le~F\`evre}}, \bibinfo
  {author} {\bibfnamefont {F.}~\bibnamefont {Bertran}}, \bibinfo {author}
  {\bibfnamefont {E.}~\bibnamefont {Giannini}}, \bibinfo {author}
  {\bibfnamefont {M.}~\bibnamefont {Grioni}}, \ and\ \bibinfo {author}
  {\bibfnamefont {V.}~\bibnamefont {Brouet}},\ }\href {\doibase
  10.1103/PhysRevLett.111.217002} {\bibfield  {journal} {\bibinfo  {journal}
  {Phys. Rev. Lett.}\ }\textbf {\bibinfo {volume} {111}},\ \bibinfo {pages}
  {217002} (\bibinfo {year} {2013})}\BibitemShut {NoStop}%
\bibitem [{\citenamefont {Uchida}\ \emph {et~al.}(2011)\citenamefont {Uchida},
  \citenamefont {Ishizaka}, \citenamefont {Hansmann}, \citenamefont {Kaneko},
  \citenamefont {Ishida}, \citenamefont {Yang}, \citenamefont {Kumai},
  \citenamefont {Toschi}, \citenamefont {Onose}, \citenamefont {Arita},
  \citenamefont {Held}, \citenamefont {Andersen}, \citenamefont {Shin},\ and\
  \citenamefont {Tokura}}]{PhysRevLett.106.027001}%
  \BibitemOpen
  \bibfield  {author} {\bibinfo {author} {\bibfnamefont {M.}~\bibnamefont
  {Uchida}}, \bibinfo {author} {\bibfnamefont {K.}~\bibnamefont {Ishizaka}},
  \bibinfo {author} {\bibfnamefont {P.}~\bibnamefont {Hansmann}}, \bibinfo
  {author} {\bibfnamefont {Y.}~\bibnamefont {Kaneko}}, \bibinfo {author}
  {\bibfnamefont {Y.}~\bibnamefont {Ishida}}, \bibinfo {author} {\bibfnamefont
  {X.}~\bibnamefont {Yang}}, \bibinfo {author} {\bibfnamefont {R.}~\bibnamefont
  {Kumai}}, \bibinfo {author} {\bibfnamefont {A.}~\bibnamefont {Toschi}},
  \bibinfo {author} {\bibfnamefont {Y.}~\bibnamefont {Onose}}, \bibinfo
  {author} {\bibfnamefont {R.}~\bibnamefont {Arita}}, \bibinfo {author}
  {\bibfnamefont {K.}~\bibnamefont {Held}}, \bibinfo {author} {\bibfnamefont
  {O.~K.}\ \bibnamefont {Andersen}}, \bibinfo {author} {\bibfnamefont
  {S.}~\bibnamefont {Shin}}, \ and\ \bibinfo {author} {\bibfnamefont
  {Y.}~\bibnamefont {Tokura}},\ }\href {\doibase
  10.1103/PhysRevLett.106.027001} {\bibfield  {journal} {\bibinfo  {journal}
  {Phys. Rev. Lett.}\ }\textbf {\bibinfo {volume} {106}},\ \bibinfo {pages}
  {027001} (\bibinfo {year} {2011})}\BibitemShut {NoStop}%
\bibitem [{\citenamefont {Kim}\ \emph {et~al.}(2014)\citenamefont {Kim},
  \citenamefont {Krupin}, \citenamefont {Denlinger}, \citenamefont {Bostwick},
  \citenamefont {Rotenberg}, \citenamefont {Zhao}, \citenamefont {Mitchell},
  \citenamefont {Allen},\ and\ \citenamefont {Kim}}]{Kim187}%
  \BibitemOpen
  \bibfield  {author} {\bibinfo {author} {\bibfnamefont {Y.~K.}\ \bibnamefont
  {Kim}}, \bibinfo {author} {\bibfnamefont {O.}~\bibnamefont {Krupin}},
  \bibinfo {author} {\bibfnamefont {J.~D.}\ \bibnamefont {Denlinger}}, \bibinfo
  {author} {\bibfnamefont {A.}~\bibnamefont {Bostwick}}, \bibinfo {author}
  {\bibfnamefont {E.}~\bibnamefont {Rotenberg}}, \bibinfo {author}
  {\bibfnamefont {Q.}~\bibnamefont {Zhao}}, \bibinfo {author} {\bibfnamefont
  {J.~F.}\ \bibnamefont {Mitchell}}, \bibinfo {author} {\bibfnamefont {J.~W.}\
  \bibnamefont {Allen}}, \ and\ \bibinfo {author} {\bibfnamefont {B.~J.}\
  \bibnamefont {Kim}},\ }\href {\doibase 10.1126/science.1251151} {\bibfield
  {journal} {\bibinfo  {journal} {Science}\ }\textbf {\bibinfo {volume}
  {345}},\ \bibinfo {pages} {187} (\bibinfo {year} {2014})}\BibitemShut
  {NoStop}%
\bibitem [{\citenamefont {Inoue}\ \emph {et~al.}(1998)\citenamefont {Inoue},
  \citenamefont {Goto}, \citenamefont {Makino}, \citenamefont {Hussey},\ and\
  \citenamefont {Ishikawa}}]{PhysRevB.58.4372}%
  \BibitemOpen
  \bibfield  {author} {\bibinfo {author} {\bibfnamefont {I.~H.}\ \bibnamefont
  {Inoue}}, \bibinfo {author} {\bibfnamefont {O.}~\bibnamefont {Goto}},
  \bibinfo {author} {\bibfnamefont {H.}~\bibnamefont {Makino}}, \bibinfo
  {author} {\bibfnamefont {N.~E.}\ \bibnamefont {Hussey}}, \ and\ \bibinfo
  {author} {\bibfnamefont {M.}~\bibnamefont {Ishikawa}},\ }\href {\doibase
  10.1103/PhysRevB.58.4372} {\bibfield  {journal} {\bibinfo  {journal} {Phys.
  Rev. B}\ }\textbf {\bibinfo {volume} {58}},\ \bibinfo {pages} {4372}
  (\bibinfo {year} {1998})}\BibitemShut {NoStop}%
\bibitem [{\citenamefont {Mo}\ \emph {et~al.}(2003)\citenamefont {Mo},
  \citenamefont {Denlinger}, \citenamefont {Kim}, \citenamefont {Park},
  \citenamefont {Allen}, \citenamefont {Sekiyama}, \citenamefont {Yamasaki},
  \citenamefont {Kadono}, \citenamefont {Suga}, \citenamefont {Saitoh},
  \citenamefont {Muro}, \citenamefont {Metcalf}, \citenamefont {Keller},
  \citenamefont {Held}, \citenamefont {Eyert}, \citenamefont {Anisimov},\ and\
  \citenamefont {Vollhardt}}]{Mo2003}%
  \BibitemOpen
  \bibfield  {author} {\bibinfo {author} {\bibfnamefont {S.-K.}\ \bibnamefont
  {Mo}}, \bibinfo {author} {\bibfnamefont {J.~D.}\ \bibnamefont {Denlinger}},
  \bibinfo {author} {\bibfnamefont {H.-D.}\ \bibnamefont {Kim}}, \bibinfo
  {author} {\bibfnamefont {J.-H.}\ \bibnamefont {Park}}, \bibinfo {author}
  {\bibfnamefont {J.~W.}\ \bibnamefont {Allen}}, \bibinfo {author}
  {\bibfnamefont {A.}~\bibnamefont {Sekiyama}}, \bibinfo {author}
  {\bibfnamefont {A.}~\bibnamefont {Yamasaki}}, \bibinfo {author}
  {\bibfnamefont {K.}~\bibnamefont {Kadono}}, \bibinfo {author} {\bibfnamefont
  {S.}~\bibnamefont {Suga}}, \bibinfo {author} {\bibfnamefont {Y.}~\bibnamefont
  {Saitoh}}, \bibinfo {author} {\bibfnamefont {T.}~\bibnamefont {Muro}},
  \bibinfo {author} {\bibfnamefont {P.}~\bibnamefont {Metcalf}}, \bibinfo
  {author} {\bibfnamefont {G.}~\bibnamefont {Keller}}, \bibinfo {author}
  {\bibfnamefont {K.}~\bibnamefont {Held}}, \bibinfo {author} {\bibfnamefont
  {V.}~\bibnamefont {Eyert}}, \bibinfo {author} {\bibfnamefont {V.~I.}\
  \bibnamefont {Anisimov}}, \ and\ \bibinfo {author} {\bibfnamefont
  {D.}~\bibnamefont {Vollhardt}},\ }\href@noop {} {\bibfield  {journal}
  {\bibinfo  {journal} {Phys. Rev. Lett.}\ }\textbf {\bibinfo {volume} {90}},\
  \bibinfo {pages} {186403} (\bibinfo {year} {2003})}\BibitemShut {NoStop}%
\bibitem [{\citenamefont {Nekrasov}\ \emph {et~al.}(2006)\citenamefont
  {Nekrasov}, \citenamefont {Held}, \citenamefont {Keller}, \citenamefont
  {Kondakov}, \citenamefont {Pruschke}, \citenamefont {Kollar}, \citenamefont
  {Andersen}, \citenamefont {Anisimov},\ and\ \citenamefont
  {Vollhardt}}]{Nekrasov2006}%
  \BibitemOpen
  \bibfield  {author} {\bibinfo {author} {\bibfnamefont {I.~A.}\ \bibnamefont
  {Nekrasov}}, \bibinfo {author} {\bibfnamefont {K.}~\bibnamefont {Held}},
  \bibinfo {author} {\bibfnamefont {G.}~\bibnamefont {Keller}}, \bibinfo
  {author} {\bibfnamefont {D.~E.}\ \bibnamefont {Kondakov}}, \bibinfo {author}
  {\bibfnamefont {T.}~\bibnamefont {Pruschke}}, \bibinfo {author}
  {\bibfnamefont {M.}~\bibnamefont {Kollar}}, \bibinfo {author} {\bibfnamefont
  {O.~K.}\ \bibnamefont {Andersen}}, \bibinfo {author} {\bibfnamefont {V.~I.}\
  \bibnamefont {Anisimov}}, \ and\ \bibinfo {author} {\bibfnamefont
  {D.}~\bibnamefont {Vollhardt}},\ }\href {\doibase 10.1103/PhysRevB.73.155112}
  {\bibfield  {journal} {\bibinfo  {journal} {Phys. Rev. B}\ }\textbf {\bibinfo
  {volume} {73}},\ \bibinfo {pages} {155112} (\bibinfo {year}
  {2006})}\BibitemShut {NoStop}%
\bibitem [{\citenamefont {Byczuk}\ \emph {et~al.}(2007)\citenamefont {Byczuk},
  \citenamefont {Kollar}, \citenamefont {Held}, \citenamefont {Yang},
  \citenamefont {Nekrasov}, \citenamefont {Pruschke},\ and\ \citenamefont
  {Vollhardt}}]{Byczuk2007}%
  \BibitemOpen
  \bibfield  {author} {\bibinfo {author} {\bibfnamefont {K.}~\bibnamefont
  {Byczuk}}, \bibinfo {author} {\bibfnamefont {M.}~\bibnamefont {Kollar}},
  \bibinfo {author} {\bibfnamefont {K.}~\bibnamefont {Held}}, \bibinfo {author}
  {\bibfnamefont {Y.~F.}\ \bibnamefont {Yang}}, \bibinfo {author}
  {\bibfnamefont {I.~A.}\ \bibnamefont {Nekrasov}}, \bibinfo {author}
  {\bibfnamefont {T.}~\bibnamefont {Pruschke}}, \ and\ \bibinfo {author}
  {\bibfnamefont {D.}~\bibnamefont {Vollhardt}},\ }\href {doi:10.1038/nphys538}
  {\bibfield  {journal} {\bibinfo  {journal} {Nat. Phys.}\ }\textbf {\bibinfo
  {volume} {3}},\ \bibinfo {pages} {168} (\bibinfo {year} {2007})}\BibitemShut
  {NoStop}%
\bibitem [{\citenamefont {Held}\ \emph {et~al.}(2013)\citenamefont {Held},
  \citenamefont {Peters},\ and\ \citenamefont {Toschi}}]{Held2013}%
  \BibitemOpen
  \bibfield  {author} {\bibinfo {author} {\bibfnamefont {K.}~\bibnamefont
  {Held}}, \bibinfo {author} {\bibfnamefont {R.}~\bibnamefont {Peters}}, \ and\
  \bibinfo {author} {\bibfnamefont {A.}~\bibnamefont {Toschi}},\ }\href
  {\doibase 10.1103/PhysRevLett.110.246402} {\bibfield  {journal} {\bibinfo
  {journal} {Phys. Rev. Lett.}\ }\textbf {\bibinfo {volume} {110}},\ \bibinfo
  {pages} {246402} (\bibinfo {year} {2013})}\BibitemShut {NoStop}%
\bibitem [{\citenamefont {Yoshimatsu}\ \emph {et~al.}(2010)\citenamefont
  {Yoshimatsu}, \citenamefont {Okabe}, \citenamefont {Kumigashira},
  \citenamefont {Okamoto}, \citenamefont {Aizaki}, \citenamefont {Fujimori},\
  and\ \citenamefont {Oshima}}]{PhysRevLett.104.147601}%
  \BibitemOpen
  \bibfield  {author} {\bibinfo {author} {\bibfnamefont {K.}~\bibnamefont
  {Yoshimatsu}}, \bibinfo {author} {\bibfnamefont {T.}~\bibnamefont {Okabe}},
  \bibinfo {author} {\bibfnamefont {H.}~\bibnamefont {Kumigashira}}, \bibinfo
  {author} {\bibfnamefont {S.}~\bibnamefont {Okamoto}}, \bibinfo {author}
  {\bibfnamefont {S.}~\bibnamefont {Aizaki}}, \bibinfo {author} {\bibfnamefont
  {A.}~\bibnamefont {Fujimori}}, \ and\ \bibinfo {author} {\bibfnamefont
  {M.}~\bibnamefont {Oshima}},\ }\href {\doibase
  10.1103/PhysRevLett.104.147601} {\bibfield  {journal} {\bibinfo  {journal}
  {Phys. Rev. Lett.}\ }\textbf {\bibinfo {volume} {104}},\ \bibinfo {pages}
  {147601} (\bibinfo {year} {2010})}\BibitemShut {NoStop}%
\bibitem [{\citenamefont {Kobayashi}\ \emph {et~al.}(2017)\citenamefont
  {Kobayashi}, \citenamefont {Yoshimatsu}, \citenamefont {Mitsuhashi},
  \citenamefont {Kitamura}, \citenamefont {Sakai}, \citenamefont {Yukawa},
  \citenamefont {Minohara}, \citenamefont {Fujimori}, \citenamefont {Horiba},\
  and\ \citenamefont {Kumigashira}}]{Kobayashi2017}%
  \BibitemOpen
  \bibfield  {author} {\bibinfo {author} {\bibfnamefont {M.}~\bibnamefont
  {Kobayashi}}, \bibinfo {author} {\bibfnamefont {K.}~\bibnamefont
  {Yoshimatsu}}, \bibinfo {author} {\bibfnamefont {T.}~\bibnamefont
  {Mitsuhashi}}, \bibinfo {author} {\bibfnamefont {M.}~\bibnamefont
  {Kitamura}}, \bibinfo {author} {\bibfnamefont {E.}~\bibnamefont {Sakai}},
  \bibinfo {author} {\bibfnamefont {R.}~\bibnamefont {Yukawa}}, \bibinfo
  {author} {\bibfnamefont {M.}~\bibnamefont {Minohara}}, \bibinfo {author}
  {\bibfnamefont {A.}~\bibnamefont {Fujimori}}, \bibinfo {author}
  {\bibfnamefont {K.}~\bibnamefont {Horiba}}, \ and\ \bibinfo {author}
  {\bibfnamefont {H.}~\bibnamefont {Kumigashira}},\ }\href {\doibase
  10.1038/s41598-017-16666-x} {\bibfield  {journal} {\bibinfo  {journal}
  {Scientific Reports}\ }\textbf {\bibinfo {volume} {7}},\ \bibinfo {pages}
  {16621} (\bibinfo {year} {2017})}\BibitemShut {NoStop}%
\bibitem [{\citenamefont {Zhong}\ \emph {et~al.}(2015)\citenamefont {Zhong},
  \citenamefont {Wallerberger}, \citenamefont {Tomczak}, \citenamefont
  {Taranto}, \citenamefont {Parragh}, \citenamefont {Toschi}, \citenamefont
  {Sangiovanni},\ and\ \citenamefont {Held}}]{PhysRevLett.114.246401}%
  \BibitemOpen
  \bibfield  {author} {\bibinfo {author} {\bibfnamefont {Z.}~\bibnamefont
  {Zhong}}, \bibinfo {author} {\bibfnamefont {M.}~\bibnamefont {Wallerberger}},
  \bibinfo {author} {\bibfnamefont {J.~M.}\ \bibnamefont {Tomczak}}, \bibinfo
  {author} {\bibfnamefont {C.}~\bibnamefont {Taranto}}, \bibinfo {author}
  {\bibfnamefont {N.}~\bibnamefont {Parragh}}, \bibinfo {author} {\bibfnamefont
  {A.}~\bibnamefont {Toschi}}, \bibinfo {author} {\bibfnamefont
  {G.}~\bibnamefont {Sangiovanni}}, \ and\ \bibinfo {author} {\bibfnamefont
  {K.}~\bibnamefont {Held}},\ }\href {\doibase 10.1103/PhysRevLett.114.246401}
  {\bibfield  {journal} {\bibinfo  {journal} {Phys. Rev. Lett.}\ }\textbf
  {\bibinfo {volume} {114}},\ \bibinfo {pages} {246401} (\bibinfo {year}
  {2015})}\BibitemShut {NoStop}%
\bibitem [{\citenamefont {Gu}\ \emph {et~al.}(2014)\citenamefont {Gu},
  \citenamefont {Wolf},\ and\ \citenamefont {Lu}}]{AMI.2014}%
  \BibitemOpen
  \bibfield  {author} {\bibinfo {author} {\bibfnamefont {M.}~\bibnamefont
  {Gu}}, \bibinfo {author} {\bibfnamefont {S.}~\bibnamefont {Wolf}}, \ and\
  \bibinfo {author} {\bibfnamefont {J.}~\bibnamefont {Lu}},\ }\href {\doibase
  10.1002/admi.201300126} {\bibfield  {journal} {\bibinfo  {journal} {Advanced
  Materials Interfaces}\ }\textbf {\bibinfo {volume} {1}} (\bibinfo {year}
  {2014}),\ 10.1002/admi.201300126}\BibitemShut {NoStop}%
\bibitem [{\citenamefont {Okada}\ \emph {et~al.}(2017)\citenamefont {Okada},
  \citenamefont {Shiau}, \citenamefont {Chang}, \citenamefont {Chang},
  \citenamefont {Kobayashi}, \citenamefont {Shimizu}, \citenamefont {Jeng},
  \citenamefont {Shiraki}, \citenamefont {Kumigashira}, \citenamefont {Bansil},
  \citenamefont {Lin},\ and\ \citenamefont
  {Hitosugi}}]{PhysRevLett.119.086801}%
  \BibitemOpen
  \bibfield  {author} {\bibinfo {author} {\bibfnamefont {Y.}~\bibnamefont
  {Okada}}, \bibinfo {author} {\bibfnamefont {S.-Y.}\ \bibnamefont {Shiau}},
  \bibinfo {author} {\bibfnamefont {T.-R.}\ \bibnamefont {Chang}}, \bibinfo
  {author} {\bibfnamefont {G.}~\bibnamefont {Chang}}, \bibinfo {author}
  {\bibfnamefont {M.}~\bibnamefont {Kobayashi}}, \bibinfo {author}
  {\bibfnamefont {R.}~\bibnamefont {Shimizu}}, \bibinfo {author} {\bibfnamefont
  {H.-T.}\ \bibnamefont {Jeng}}, \bibinfo {author} {\bibfnamefont
  {S.}~\bibnamefont {Shiraki}}, \bibinfo {author} {\bibfnamefont
  {H.}~\bibnamefont {Kumigashira}}, \bibinfo {author} {\bibfnamefont
  {A.}~\bibnamefont {Bansil}}, \bibinfo {author} {\bibfnamefont
  {H.}~\bibnamefont {Lin}}, \ and\ \bibinfo {author} {\bibfnamefont
  {T.}~\bibnamefont {Hitosugi}},\ }\href {\doibase
  10.1103/PhysRevLett.119.086801} {\bibfield  {journal} {\bibinfo  {journal}
  {Phys. Rev. Lett.}\ }\textbf {\bibinfo {volume} {119}},\ \bibinfo {pages}
  {086801} (\bibinfo {year} {2017})}\BibitemShut {NoStop}%
\bibitem [{\citenamefont {Gabel}\ \emph {et~al.}(2021)\citenamefont {Gabel},
  \citenamefont {Pickem}, \citenamefont {Scheiderer}, \citenamefont {Dudy},
  \citenamefont {Leikert}, \citenamefont {Fuchs}, \citenamefont {Stübinger},
  \citenamefont {Schmitt}, \citenamefont {Küspert}, \citenamefont
  {Sangiovanni}, \citenamefont {Tomczak}, \citenamefont {Held}, \citenamefont
  {Lee}, \citenamefont {Claessen},\ and\ \citenamefont {Sing}}]{Gabel2022}%
  \BibitemOpen
  \bibfield  {author} {\bibinfo {author} {\bibfnamefont {J.}~\bibnamefont
  {Gabel}}, \bibinfo {author} {\bibfnamefont {M.}~\bibnamefont {Pickem}},
  \bibinfo {author} {\bibfnamefont {P.}~\bibnamefont {Scheiderer}}, \bibinfo
  {author} {\bibfnamefont {L.}~\bibnamefont {Dudy}}, \bibinfo {author}
  {\bibfnamefont {B.}~\bibnamefont {Leikert}}, \bibinfo {author} {\bibfnamefont
  {M.}~\bibnamefont {Fuchs}}, \bibinfo {author} {\bibfnamefont
  {M.}~\bibnamefont {Stübinger}}, \bibinfo {author} {\bibfnamefont
  {M.}~\bibnamefont {Schmitt}}, \bibinfo {author} {\bibfnamefont
  {J.}~\bibnamefont {Küspert}}, \bibinfo {author} {\bibfnamefont
  {G.}~\bibnamefont {Sangiovanni}}, \bibinfo {author} {\bibfnamefont {J.~M.}\
  \bibnamefont {Tomczak}}, \bibinfo {author} {\bibfnamefont {K.}~\bibnamefont
  {Held}}, \bibinfo {author} {\bibfnamefont {T.-L.}\ \bibnamefont {Lee}},
  \bibinfo {author} {\bibfnamefont {R.}~\bibnamefont {Claessen}}, \ and\
  \bibinfo {author} {\bibfnamefont {M.}~\bibnamefont {Sing}},\ }\href {\doibase
  https://doi.org/10.1002/aelm.202101006} {\bibfield  {journal} {\bibinfo
  {journal} {Advanced Electronic Materials}\ }\textbf {\bibinfo {volume}
  {n/a}},\ \bibinfo {pages} {2101006} (\bibinfo {year} {2021})}\BibitemShut
  {NoStop}%
\bibitem [{\citenamefont {Metzner}\ and\ \citenamefont
  {Vollhardt}(1989)}]{PhysRevLett.62.324}%
  \BibitemOpen
  \bibfield  {author} {\bibinfo {author} {\bibfnamefont {W.}~\bibnamefont
  {Metzner}}\ and\ \bibinfo {author} {\bibfnamefont {D.}~\bibnamefont
  {Vollhardt}},\ }\href {\doibase 10.1103/PhysRevLett.62.324} {\bibfield
  {journal} {\bibinfo  {journal} {Phys. Rev. Lett.}\ }\textbf {\bibinfo
  {volume} {62}},\ \bibinfo {pages} {324} (\bibinfo {year} {1989})}\BibitemShut
  {NoStop}%
\bibitem [{\citenamefont {Georges}\ and\ \citenamefont
  {Kotliar}(1992)}]{PhysRevB.45.6479}%
  \BibitemOpen
  \bibfield  {author} {\bibinfo {author} {\bibfnamefont {A.}~\bibnamefont
  {Georges}}\ and\ \bibinfo {author} {\bibfnamefont {G.}~\bibnamefont
  {Kotliar}},\ }\href {\doibase 10.1103/PhysRevB.45.6479} {\bibfield  {journal}
  {\bibinfo  {journal} {Phys. Rev. B}\ }\textbf {\bibinfo {volume} {45}},\
  \bibinfo {pages} {6479} (\bibinfo {year} {1992})}\BibitemShut {NoStop}%
\bibitem [{\citenamefont {Georges}\ \emph {et~al.}(1996)\citenamefont
  {Georges}, \citenamefont {Kotliar}, \citenamefont {Krauth},\ and\
  \citenamefont {Rozenberg}}]{Georges96}%
  \BibitemOpen
  \bibfield  {author} {\bibinfo {author} {\bibfnamefont {A.}~\bibnamefont
  {Georges}}, \bibinfo {author} {\bibfnamefont {G.}~\bibnamefont {Kotliar}},
  \bibinfo {author} {\bibfnamefont {W.}~\bibnamefont {Krauth}}, \ and\ \bibinfo
  {author} {\bibfnamefont {M.~J.}\ \bibnamefont {Rozenberg}},\ }\href {\doibase
  10.1103/RevModPhys.68.13} {\bibfield  {journal} {\bibinfo  {journal} {Rev.
  Mod. Phys.}\ }\textbf {\bibinfo {volume} {68}},\ \bibinfo {pages} {13}
  (\bibinfo {year} {1996})}\BibitemShut {NoStop}%
\bibitem [{\citenamefont {Toschi}\ \emph {et~al.}(2007)\citenamefont {Toschi},
  \citenamefont {Katanin},\ and\ \citenamefont {Held}}]{toschi:045118}%
  \BibitemOpen
  \bibfield  {author} {\bibinfo {author} {\bibfnamefont {A.}~\bibnamefont
  {Toschi}}, \bibinfo {author} {\bibfnamefont {A.~A.}\ \bibnamefont {Katanin}},
  \ and\ \bibinfo {author} {\bibfnamefont {K.}~\bibnamefont {Held}},\ }\href
  {\doibase 10.1103/PhysRevB.75.045118} {\bibfield  {journal} {\bibinfo
  {journal} {Phys. Rev. B}\ }\textbf {\bibinfo {volume} {75}},\ \bibinfo {eid}
  {045118} (\bibinfo {year} {2007})}\BibitemShut {NoStop}%
\bibitem [{\citenamefont {Katanin}\ \emph {et~al.}(2009)\citenamefont
  {Katanin}, \citenamefont {Toschi},\ and\ \citenamefont {Held}}]{Katanin2009}%
  \BibitemOpen
  \bibfield  {author} {\bibinfo {author} {\bibfnamefont {A.~A.}\ \bibnamefont
  {Katanin}}, \bibinfo {author} {\bibfnamefont {A.}~\bibnamefont {Toschi}}, \
  and\ \bibinfo {author} {\bibfnamefont {K.}~\bibnamefont {Held}},\ }\href
  {\doibase 10.1103/PhysRevB.80.075104} {\bibfield  {journal} {\bibinfo
  {journal} {Phys. Rev. B}\ }\textbf {\bibinfo {volume} {80}},\ \bibinfo
  {pages} {075104} (\bibinfo {year} {2009})}\BibitemShut {NoStop}%
\bibitem [{\citenamefont {Galler}\ \emph {et~al.}(2017)\citenamefont {Galler},
  \citenamefont {Thunstr{\"o}m}, \citenamefont {Gunacker}, \citenamefont
  {Tomczak},\ and\ \citenamefont {Held}}]{Anna_ADGA}%
  \BibitemOpen
  \bibfield  {author} {\bibinfo {author} {\bibfnamefont {A.}~\bibnamefont
  {Galler}}, \bibinfo {author} {\bibfnamefont {P.}~\bibnamefont
  {Thunstr{\"o}m}}, \bibinfo {author} {\bibfnamefont {P.}~\bibnamefont
  {Gunacker}}, \bibinfo {author} {\bibfnamefont {J.~M.}\ \bibnamefont
  {Tomczak}}, \ and\ \bibinfo {author} {\bibfnamefont {K.}~\bibnamefont
  {Held}},\ }\href {\doibase 10.1103/PhysRevB.95.115107} {\bibfield  {journal}
  {\bibinfo  {journal} {Phys. Rev. B}\ }\textbf {\bibinfo {volume} {95}},\
  \bibinfo {pages} {115107} (\bibinfo {year} {2017})}\BibitemShut {NoStop}%
\bibitem [{\citenamefont {Rohringer}\ \emph {et~al.}(2018)\citenamefont
  {Rohringer}, \citenamefont {Hafermann}, \citenamefont {Toschi}, \citenamefont
  {Katanin}, \citenamefont {Antipov}, \citenamefont {Katsnelson}, \citenamefont
  {Lichtenstein}, \citenamefont {Rubtsov},\ and\ \citenamefont
  {Held}}]{RevModPhys.90.025003}%
  \BibitemOpen
  \bibfield  {author} {\bibinfo {author} {\bibfnamefont {G.}~\bibnamefont
  {Rohringer}}, \bibinfo {author} {\bibfnamefont {H.}~\bibnamefont
  {Hafermann}}, \bibinfo {author} {\bibfnamefont {A.}~\bibnamefont {Toschi}},
  \bibinfo {author} {\bibfnamefont {A.~A.}\ \bibnamefont {Katanin}}, \bibinfo
  {author} {\bibfnamefont {A.~E.}\ \bibnamefont {Antipov}}, \bibinfo {author}
  {\bibfnamefont {M.~I.}\ \bibnamefont {Katsnelson}}, \bibinfo {author}
  {\bibfnamefont {A.~I.}\ \bibnamefont {Lichtenstein}}, \bibinfo {author}
  {\bibfnamefont {A.~N.}\ \bibnamefont {Rubtsov}}, \ and\ \bibinfo {author}
  {\bibfnamefont {K.}~\bibnamefont {Held}},\ }\href {\doibase
  10.1103/RevModPhys.90.025003} {\bibfield  {journal} {\bibinfo  {journal}
  {Rev. Mod. Phys.}\ }\textbf {\bibinfo {volume} {90}},\ \bibinfo {pages}
  {025003} (\bibinfo {year} {2018})}\BibitemShut {NoStop}%
\bibitem [{\citenamefont {Godby}\ \emph {et~al.}(1988)\citenamefont {Godby},
  \citenamefont {Schl\"uter},\ and\ \citenamefont {Sham}}]{Godby88}%
  \BibitemOpen
  \bibfield  {author} {\bibinfo {author} {\bibfnamefont {R.~W.}\ \bibnamefont
  {Godby}}, \bibinfo {author} {\bibfnamefont {M.}~\bibnamefont {Schl\"uter}}, \
  and\ \bibinfo {author} {\bibfnamefont {L.~J.}\ \bibnamefont {Sham}},\ }\href
  {\doibase 10.1103/PhysRevB.37.10159} {\bibfield  {journal} {\bibinfo
  {journal} {Phys. Rev. B}\ }\textbf {\bibinfo {volume} {37}},\ \bibinfo
  {pages} {10159} (\bibinfo {year} {1988})}\BibitemShut {NoStop}%
\bibitem [{\citenamefont {Miyake}\ \emph {et~al.}(2013)\citenamefont {Miyake},
  \citenamefont {Martins}, \citenamefont {Sakuma},\ and\ \citenamefont
  {Aryasetiawan}}]{PhysRevB.87.115110}%
  \BibitemOpen
  \bibfield  {author} {\bibinfo {author} {\bibfnamefont {T.}~\bibnamefont
  {Miyake}}, \bibinfo {author} {\bibfnamefont {C.}~\bibnamefont {Martins}},
  \bibinfo {author} {\bibfnamefont {R.}~\bibnamefont {Sakuma}}, \ and\ \bibinfo
  {author} {\bibfnamefont {F.}~\bibnamefont {Aryasetiawan}},\ }\href {\doibase
  10.1103/PhysRevB.87.115110} {\bibfield  {journal} {\bibinfo  {journal} {Phys.
  Rev. B}\ }\textbf {\bibinfo {volume} {87}},\ \bibinfo {pages} {115110}
  (\bibinfo {year} {2013})}\BibitemShut {NoStop}%
\bibitem [{\citenamefont {Tomczak}\ \emph {et~al.}(2014)\citenamefont
  {Tomczak}, \citenamefont {Casula}, \citenamefont {Miyake},\ and\
  \citenamefont {Biermann}}]{Tomczak14}%
  \BibitemOpen
  \bibfield  {author} {\bibinfo {author} {\bibfnamefont {J.~M.}\ \bibnamefont
  {Tomczak}}, \bibinfo {author} {\bibfnamefont {M.}~\bibnamefont {Casula}},
  \bibinfo {author} {\bibfnamefont {T.}~\bibnamefont {Miyake}}, \ and\ \bibinfo
  {author} {\bibfnamefont {S.}~\bibnamefont {Biermann}},\ }\href {\doibase
  10.1103/PhysRevB.90.165138} {\bibfield  {journal} {\bibinfo  {journal} {Phys.
  Rev. B}\ }\textbf {\bibinfo {volume} {90}},\ \bibinfo {pages} {165138}
  (\bibinfo {year} {2014})}\BibitemShut {NoStop}%
\bibitem [{\citenamefont {Blaha}\ \emph {et~al.}(2001)\citenamefont {Blaha},
  \citenamefont {Schwarz}, \citenamefont {Madsen}, \citenamefont {Kvasnicka},\
  and\ \citenamefont {Luitz}}]{wien2k}%
  \BibitemOpen
  \bibfield  {author} {\bibinfo {author} {\bibfnamefont {P.}~\bibnamefont
  {Blaha}}, \bibinfo {author} {\bibfnamefont {K.}~\bibnamefont {Schwarz}},
  \bibinfo {author} {\bibfnamefont {G.-K.-H.}\ \bibnamefont {Madsen}}, \bibinfo
  {author} {\bibfnamefont {D.}~\bibnamefont {Kvasnicka}}, \ and\ \bibinfo
  {author} {\bibfnamefont {J.}~\bibnamefont {Luitz}},\ }\href@noop {}
  {\bibfield  {journal} {\bibinfo  {journal} {Vienna University of Technology,
  Austria}\ } (\bibinfo {year} {2001})},\ \bibinfo {note} {{ISBN}
  3-9501031-1-2}\BibitemShut {NoStop}%
\bibitem [{\citenamefont {Blaha}\ \emph {et~al.}(2020)\citenamefont {Blaha},
  \citenamefont {Schwarz}, \citenamefont {Tran}, \citenamefont {Laskowski},
  \citenamefont {Madsen},\ and\ \citenamefont {Marks}}]{doi:10.1063/1.5143061}%
  \BibitemOpen
  \bibfield  {author} {\bibinfo {author} {\bibfnamefont {P.}~\bibnamefont
  {Blaha}}, \bibinfo {author} {\bibfnamefont {K.}~\bibnamefont {Schwarz}},
  \bibinfo {author} {\bibfnamefont {F.}~\bibnamefont {Tran}}, \bibinfo {author}
  {\bibfnamefont {R.}~\bibnamefont {Laskowski}}, \bibinfo {author}
  {\bibfnamefont {G.~K.~H.}\ \bibnamefont {Madsen}}, \ and\ \bibinfo {author}
  {\bibfnamefont {L.~D.}\ \bibnamefont {Marks}},\ }\href {\doibase
  10.1063/1.5143061} {\bibfield  {journal} {\bibinfo  {journal} {The Journal of
  Chemical Physics}\ }\textbf {\bibinfo {volume} {152}},\ \bibinfo {pages}
  {074101} (\bibinfo {year} {2020})}\BibitemShut {NoStop}%
\bibitem [{\citenamefont {Perdew}\ \emph {et~al.}(1996)\citenamefont {Perdew},
  \citenamefont {Burke},\ and\ \citenamefont
  {Ernzerhof}}]{PhysRevLett.77.3865}%
  \BibitemOpen
  \bibfield  {author} {\bibinfo {author} {\bibfnamefont {J.~P.}\ \bibnamefont
  {Perdew}}, \bibinfo {author} {\bibfnamefont {K.}~\bibnamefont {Burke}}, \
  and\ \bibinfo {author} {\bibfnamefont {M.}~\bibnamefont {Ernzerhof}},\ }\href
  {\doibase 10.1103/PhysRevLett.77.3865} {\bibfield  {journal} {\bibinfo
  {journal} {Phys. Rev. Lett.}\ }\textbf {\bibinfo {volume} {77}},\ \bibinfo
  {pages} {3865} (\bibinfo {year} {1996})}\BibitemShut {NoStop}%
\bibitem [{\citenamefont {Wang}\ \emph {et~al.}(2019)\citenamefont {Wang},
  \citenamefont {Zhang}, \citenamefont {Deepak}, \citenamefont {Chen},
  \citenamefont {Fouchet}, \citenamefont {Duan}, \citenamefont {Hilliard},
  \citenamefont {Kentsch}, \citenamefont {Chen}, \citenamefont {Zeng},
  \citenamefont {Gao}, \citenamefont {Zeng}, \citenamefont {Helm},
  \citenamefont {Prellier},\ and\ \citenamefont
  {Zhou}}]{PhysRevMaterials.3.115001}%
  \BibitemOpen
  \bibfield  {author} {\bibinfo {author} {\bibfnamefont {C.}~\bibnamefont
  {Wang}}, \bibinfo {author} {\bibfnamefont {H.}~\bibnamefont {Zhang}},
  \bibinfo {author} {\bibfnamefont {K.}~\bibnamefont {Deepak}}, \bibinfo
  {author} {\bibfnamefont {C.}~\bibnamefont {Chen}}, \bibinfo {author}
  {\bibfnamefont {A.}~\bibnamefont {Fouchet}}, \bibinfo {author} {\bibfnamefont
  {J.}~\bibnamefont {Duan}}, \bibinfo {author} {\bibfnamefont {D.}~\bibnamefont
  {Hilliard}}, \bibinfo {author} {\bibfnamefont {U.}~\bibnamefont {Kentsch}},
  \bibinfo {author} {\bibfnamefont {D.}~\bibnamefont {Chen}}, \bibinfo {author}
  {\bibfnamefont {M.}~\bibnamefont {Zeng}}, \bibinfo {author} {\bibfnamefont
  {X.}~\bibnamefont {Gao}}, \bibinfo {author} {\bibfnamefont {Y.-J.}\
  \bibnamefont {Zeng}}, \bibinfo {author} {\bibfnamefont {M.}~\bibnamefont
  {Helm}}, \bibinfo {author} {\bibfnamefont {W.}~\bibnamefont {Prellier}}, \
  and\ \bibinfo {author} {\bibfnamefont {S.}~\bibnamefont {Zhou}},\ }\href
  {\doibase 10.1103/PhysRevMaterials.3.115001} {\bibfield  {journal} {\bibinfo
  {journal} {Phys. Rev. Materials}\ }\textbf {\bibinfo {volume} {3}},\ \bibinfo
  {pages} {115001} (\bibinfo {year} {2019})}\BibitemShut {NoStop}%
\bibitem [{\citenamefont {Kunes}\ \emph {et~al.}(2010)\citenamefont {Kunes},
  \citenamefont {Arita}, \citenamefont {Wissgott}, \citenamefont {Toschi},
  \citenamefont {Ikeda},\ and\ \citenamefont {Held}}]{wien2wannier}%
  \BibitemOpen
  \bibfield  {author} {\bibinfo {author} {\bibfnamefont {J.}~\bibnamefont
  {Kunes}}, \bibinfo {author} {\bibfnamefont {R.}~\bibnamefont {Arita}},
  \bibinfo {author} {\bibfnamefont {P.}~\bibnamefont {Wissgott}}, \bibinfo
  {author} {\bibfnamefont {A.}~\bibnamefont {Toschi}}, \bibinfo {author}
  {\bibfnamefont {H.}~\bibnamefont {Ikeda}}, \ and\ \bibinfo {author}
  {\bibfnamefont {K.}~\bibnamefont {Held}},\ }\href {\doibase
  http://dx.doi.org/10.1016/j.cpc.2010.08.005} {\bibfield  {journal} {\bibinfo
  {journal} {Computer Physics Communications}\ }\textbf {\bibinfo {volume}
  {181}},\ \bibinfo {pages} {1888 } (\bibinfo {year} {2010})}\BibitemShut
  {NoStop}%
\bibitem [{\citenamefont {Mostofi}\ \emph {et~al.}(2008)\citenamefont
  {Mostofi}, \citenamefont {Yates}, \citenamefont {Lee}, \citenamefont {Souza},
  \citenamefont {Vanderbilt},\ and\ \citenamefont {Marzari}}]{wannier90}%
  \BibitemOpen
  \bibfield  {author} {\bibinfo {author} {\bibfnamefont {A.~A.}\ \bibnamefont
  {Mostofi}}, \bibinfo {author} {\bibfnamefont {J.~R.}\ \bibnamefont {Yates}},
  \bibinfo {author} {\bibfnamefont {Y.-S.}\ \bibnamefont {Lee}}, \bibinfo
  {author} {\bibfnamefont {I.}~\bibnamefont {Souza}}, \bibinfo {author}
  {\bibfnamefont {D.}~\bibnamefont {Vanderbilt}}, \ and\ \bibinfo {author}
  {\bibfnamefont {N.}~\bibnamefont {Marzari}},\ }\href {\doibase
  https://doi.org/10.1016/j.cpc.2007.11.016} {\bibfield  {journal} {\bibinfo
  {journal} {Computer Physics Communications}\ }\textbf {\bibinfo {volume}
  {178}},\ \bibinfo {pages} {685 } (\bibinfo {year} {2008})}\BibitemShut
  {NoStop}%
\bibitem [{\citenamefont {Held}(2007)}]{doi:10.1080/00018730701619647}%
  \BibitemOpen
  \bibfield  {author} {\bibinfo {author} {\bibfnamefont {K.}~\bibnamefont
  {Held}},\ }\href {\doibase 10.1080/00018730701619647} {\bibfield  {journal}
  {\bibinfo  {journal} {Advances in Physics}\ }\textbf {\bibinfo {volume}
  {56}},\ \bibinfo {pages} {829} (\bibinfo {year} {2007})}\BibitemShut
  {NoStop}%
\bibitem [{\citenamefont {Kaufmann}(2019)}]{ana_cont}%
  \BibitemOpen
  \bibfield  {author} {\bibinfo {author} {\bibfnamefont {J.}~\bibnamefont
  {Kaufmann}},\ }\href {$github.com/josefkaufmann/ana_cont$} {\bibfield
  {journal} {\bibinfo  {journal} {$github.com/josefkaufmann/ana\_cont$}\ }
  (\bibinfo {year} {2019})}\BibitemShut {NoStop}%
\bibitem [{\citenamefont {Kaufmann}\ and\ \citenamefont
  {Held}(2021)}]{kaufmann2021anacont}%
  \BibitemOpen
  \bibfield  {author} {\bibinfo {author} {\bibfnamefont {J.}~\bibnamefont
  {Kaufmann}}\ and\ \bibinfo {author} {\bibfnamefont {K.}~\bibnamefont
  {Held}},\ }\href@noop {} {\  (\bibinfo {year} {2021})},\ \Eprint
  {http://arxiv.org/abs/2105.11211} {arXiv:2105.11211 [cond-mat.str-el]}
  \BibitemShut {NoStop}%
\bibitem [{\citenamefont {Tomczak}\ \emph {et~al.}(2017)\citenamefont
  {Tomczak}, \citenamefont {Liu}, \citenamefont {Toschi}, \citenamefont
  {Kresse},\ and\ \citenamefont {Held}}]{Tomczak2017review}%
  \BibitemOpen
  \bibfield  {author} {\bibinfo {author} {\bibfnamefont {J.~M.}\ \bibnamefont
  {Tomczak}}, \bibinfo {author} {\bibfnamefont {P.}~\bibnamefont {Liu}},
  \bibinfo {author} {\bibfnamefont {A.}~\bibnamefont {Toschi}}, \bibinfo
  {author} {\bibfnamefont {G.}~\bibnamefont {Kresse}}, \ and\ \bibinfo {author}
  {\bibfnamefont {K.}~\bibnamefont {Held}},\ }\href {\doibase
  10.1140/epjst/e2017-70053-1} {\bibfield  {journal} {\bibinfo  {journal} {The
  European Physical Journal Special Topics}\ }\textbf {\bibinfo {volume}
  {226}},\ \bibinfo {pages} {2565} (\bibinfo {year} {2017})}\BibitemShut
  {NoStop}%
\bibitem [{\citenamefont {Galler}\ \emph {et~al.}(2018)\citenamefont {Galler},
  \citenamefont {Kaufmann}, \citenamefont {Gunacker}, \citenamefont {Pickem},
  \citenamefont {Thunstr{\"o}m}, \citenamefont {Tomczak},\ and\ \citenamefont
  {Held}}]{JPSJ_ADGA}%
  \BibitemOpen
  \bibfield  {author} {\bibinfo {author} {\bibfnamefont {A.}~\bibnamefont
  {Galler}}, \bibinfo {author} {\bibfnamefont {J.}~\bibnamefont {Kaufmann}},
  \bibinfo {author} {\bibfnamefont {P.}~\bibnamefont {Gunacker}}, \bibinfo
  {author} {\bibfnamefont {M.}~\bibnamefont {Pickem}}, \bibinfo {author}
  {\bibfnamefont {P.}~\bibnamefont {Thunstr{\"o}m}}, \bibinfo {author}
  {\bibfnamefont {J.~M.}\ \bibnamefont {Tomczak}}, \ and\ \bibinfo {author}
  {\bibfnamefont {K.}~\bibnamefont {Held}},\ }\href {\doibase
  10.7566/JPSJ.87.041004} {\bibfield  {journal} {\bibinfo  {journal} {Journal
  of the Physical Society of Japan}\ }\textbf {\bibinfo {volume} {87}},\
  \bibinfo {pages} {041004} (\bibinfo {year} {2018})}\BibitemShut {NoStop}%
\bibitem [{\citenamefont {Galler}\ \emph {et~al.}(2019)\citenamefont {Galler},
  \citenamefont {Thunström}, \citenamefont {Kaufmann}, \citenamefont {Pickem},
  \citenamefont {Tomczak},\ and\ \citenamefont {Held}}]{CPC_ADGA}%
  \BibitemOpen
  \bibfield  {author} {\bibinfo {author} {\bibfnamefont {A.}~\bibnamefont
  {Galler}}, \bibinfo {author} {\bibfnamefont {P.}~\bibnamefont {Thunström}},
  \bibinfo {author} {\bibfnamefont {J.}~\bibnamefont {Kaufmann}}, \bibinfo
  {author} {\bibfnamefont {M.}~\bibnamefont {Pickem}}, \bibinfo {author}
  {\bibfnamefont {J.~M.}\ \bibnamefont {Tomczak}}, \ and\ \bibinfo {author}
  {\bibfnamefont {K.}~\bibnamefont {Held}},\ }\href {\doibase
  https://doi.org/10.1016/j.cpc.2019.07.012} {\bibfield  {journal} {\bibinfo
  {journal} {Computer Physics Communications}\ }\textbf {\bibinfo {volume}
  {245}},\ \bibinfo {pages} {106847} (\bibinfo {year} {2019})}\BibitemShut
  {NoStop}%
\bibitem [{\citenamefont {Kusunose}(2006)}]{Kusunose2006}%
  \BibitemOpen
  \bibfield  {author} {\bibinfo {author} {\bibfnamefont {H.}~\bibnamefont
  {Kusunose}},\ }\href {\doibase 10.1143/JPSJ.75.054713} {\bibfield  {journal}
  {\bibinfo  {journal} {J. Phys. Soc. Jpn.}\ }\textbf {\bibinfo {volume}
  {75}},\ \bibinfo {pages} {054713} (\bibinfo {year} {2006})}\BibitemShut
  {NoStop}%
\bibitem [{\citenamefont {Rubtsov}\ \emph {et~al.}(2008)\citenamefont
  {Rubtsov}, \citenamefont {Katsnelson},\ and\ \citenamefont
  {Lichtenstein}}]{PhysRevB.77.033101}%
  \BibitemOpen
  \bibfield  {author} {\bibinfo {author} {\bibfnamefont {A.~N.}\ \bibnamefont
  {Rubtsov}}, \bibinfo {author} {\bibfnamefont {M.~I.}\ \bibnamefont
  {Katsnelson}}, \ and\ \bibinfo {author} {\bibfnamefont {A.~I.}\ \bibnamefont
  {Lichtenstein}},\ }\href {\doibase 10.1103/PhysRevB.77.033101} {\bibfield
  {journal} {\bibinfo  {journal} {Phys. Rev. B}\ }\textbf {\bibinfo {volume}
  {77}},\ \bibinfo {pages} {033101} (\bibinfo {year} {2008})}\BibitemShut
  {NoStop}%
\bibitem [{\citenamefont {Rohringer}\ \emph {et~al.}(2013)\citenamefont
  {Rohringer}, \citenamefont {Toschi}, \citenamefont {Hafermann}, \citenamefont
  {Held}, \citenamefont {Anisimov},\ and\ \citenamefont
  {Katanin}}]{Rohringer2013}%
  \BibitemOpen
  \bibfield  {author} {\bibinfo {author} {\bibfnamefont {G.}~\bibnamefont
  {Rohringer}}, \bibinfo {author} {\bibfnamefont {A.}~\bibnamefont {Toschi}},
  \bibinfo {author} {\bibfnamefont {H.}~\bibnamefont {Hafermann}}, \bibinfo
  {author} {\bibfnamefont {K.}~\bibnamefont {Held}}, \bibinfo {author}
  {\bibfnamefont {V.~I.}\ \bibnamefont {Anisimov}}, \ and\ \bibinfo {author}
  {\bibfnamefont {A.~A.}\ \bibnamefont {Katanin}},\ }\href {\doibase
  10.1103/PhysRevB.88.115112} {\bibfield  {journal} {\bibinfo  {journal} {Phys.
  Rev. B}\ }\textbf {\bibinfo {volume} {88}},\ \bibinfo {pages} {115112}
  (\bibinfo {year} {2013})}\BibitemShut {NoStop}%
\bibitem [{\citenamefont {Taranto}\ \emph {et~al.}(2014)\citenamefont
  {Taranto}, \citenamefont {Andergassen}, \citenamefont {Bauer}, \citenamefont
  {Held}, \citenamefont {Katanin}, \citenamefont {Metzner}, \citenamefont
  {Rohringer},\ and\ \citenamefont {Toschi}}]{Taranto2014}%
  \BibitemOpen
  \bibfield  {author} {\bibinfo {author} {\bibfnamefont {C.}~\bibnamefont
  {Taranto}}, \bibinfo {author} {\bibfnamefont {S.}~\bibnamefont
  {Andergassen}}, \bibinfo {author} {\bibfnamefont {J.}~\bibnamefont {Bauer}},
  \bibinfo {author} {\bibfnamefont {K.}~\bibnamefont {Held}}, \bibinfo {author}
  {\bibfnamefont {A.}~\bibnamefont {Katanin}}, \bibinfo {author} {\bibfnamefont
  {W.}~\bibnamefont {Metzner}}, \bibinfo {author} {\bibfnamefont
  {G.}~\bibnamefont {Rohringer}}, \ and\ \bibinfo {author} {\bibfnamefont
  {A.}~\bibnamefont {Toschi}},\ }\href {\doibase
  10.1103/PhysRevLett.112.196402} {\bibfield  {journal} {\bibinfo  {journal}
  {Phys. Rev. Lett.}\ }\textbf {\bibinfo {volume} {112}},\ \bibinfo {pages}
  {196402} (\bibinfo {year} {2014})}\BibitemShut {NoStop}%
\bibitem [{\citenamefont {Li}(2015)}]{Li2015}%
  \BibitemOpen
  \bibfield  {author} {\bibinfo {author} {\bibfnamefont {G.}~\bibnamefont
  {Li}},\ }\href {\doibase 10.1103/PhysRevB.91.165134} {\bibfield  {journal}
  {\bibinfo  {journal} {Phys. Rev. B}\ }\textbf {\bibinfo {volume} {91}},\
  \bibinfo {pages} {165134} (\bibinfo {year} {2015})}\BibitemShut {NoStop}%
\bibitem [{\citenamefont {Parcollet}\ \emph {et~al.}(2015)\citenamefont
  {Parcollet}, \citenamefont {Ferrero}, \citenamefont {Ayral}, \citenamefont
  {Hafermann}, \citenamefont {Krivenko}, \citenamefont {Messio},\ and\
  \citenamefont {Seth}}]{TRIQS}%
  \BibitemOpen
  \bibfield  {author} {\bibinfo {author} {\bibfnamefont {O.}~\bibnamefont
  {Parcollet}}, \bibinfo {author} {\bibfnamefont {M.}~\bibnamefont {Ferrero}},
  \bibinfo {author} {\bibfnamefont {T.}~\bibnamefont {Ayral}}, \bibinfo
  {author} {\bibfnamefont {H.}~\bibnamefont {Hafermann}}, \bibinfo {author}
  {\bibfnamefont {I.}~\bibnamefont {Krivenko}}, \bibinfo {author}
  {\bibfnamefont {L.}~\bibnamefont {Messio}}, \ and\ \bibinfo {author}
  {\bibfnamefont {P.}~\bibnamefont {Seth}},\ }\href {\doibase
  http://dx.doi.org/10.1016/j.cpc.2015.04.023} {\bibfield  {journal} {\bibinfo
  {journal} {Computer Physics Communications}\ }\textbf {\bibinfo {volume}
  {196}},\ \bibinfo {pages} {398 } (\bibinfo {year} {2015})}\BibitemShut
  {NoStop}%
\bibitem [{\citenamefont {Sordi}\ \emph {et~al.}(2011)\citenamefont {Sordi},
  \citenamefont {Haule},\ and\ \citenamefont {Tremblay}}]{Sordi2011}%
  \BibitemOpen
  \bibfield  {author} {\bibinfo {author} {\bibfnamefont {G.}~\bibnamefont
  {Sordi}}, \bibinfo {author} {\bibfnamefont {K.}~\bibnamefont {Haule}}, \ and\
  \bibinfo {author} {\bibfnamefont {A.-M.~S.}\ \bibnamefont {Tremblay}},\
  }\href {\doibase 10.1103/PhysRevB.84.075161} {\bibfield  {journal} {\bibinfo
  {journal} {Phys. Rev. B}\ }\textbf {\bibinfo {volume} {84}},\ \bibinfo
  {pages} {075161} (\bibinfo {year} {2011})}\BibitemShut {NoStop}%
\bibitem [{\citenamefont {Sakai}\ \emph {et~al.}(2012)\citenamefont {Sakai},
  \citenamefont {Sangiovanni}, \citenamefont {Civelli}, \citenamefont {Motome},
  \citenamefont {Held},\ and\ \citenamefont {Imada}}]{Sakai12}%
  \BibitemOpen
  \bibfield  {author} {\bibinfo {author} {\bibfnamefont {S.}~\bibnamefont
  {Sakai}}, \bibinfo {author} {\bibfnamefont {G.}~\bibnamefont {Sangiovanni}},
  \bibinfo {author} {\bibfnamefont {M.}~\bibnamefont {Civelli}}, \bibinfo
  {author} {\bibfnamefont {Y.}~\bibnamefont {Motome}}, \bibinfo {author}
  {\bibfnamefont {K.}~\bibnamefont {Held}}, \ and\ \bibinfo {author}
  {\bibfnamefont {M.}~\bibnamefont {Imada}},\ }\href {\doibase
  10.1103/PhysRevB.85.035102} {\bibfield  {journal} {\bibinfo  {journal} {Phys.
  Rev. B}\ }\textbf {\bibinfo {volume} {85}},\ \bibinfo {pages} {035102}
  (\bibinfo {year} {2012})}\BibitemShut {NoStop}%
\bibitem [{\citenamefont {Sch\"afer}\ \emph {et~al.}(2016)\citenamefont
  {Sch\"afer}, \citenamefont {Toschi},\ and\ \citenamefont
  {Held}}]{Schaefer2015-3}%
  \BibitemOpen
  \bibfield  {author} {\bibinfo {author} {\bibfnamefont {T.}~\bibnamefont
  {Sch\"afer}}, \bibinfo {author} {\bibfnamefont {A.}~\bibnamefont {Toschi}}, \
  and\ \bibinfo {author} {\bibfnamefont {K.}~\bibnamefont {Held}},\ }\href
  {\doibase 10.1016/j.jmmm.2015.07.103} {\bibfield  {journal} {\bibinfo
  {journal} {Journal of Magnetism and Magnetic Materials}\ }\textbf {\bibinfo
  {volume} {400}},\ \bibinfo {pages} {107} (\bibinfo {year}
  {2016})}\BibitemShut {NoStop}%
\bibitem [{\citenamefont {Sch\"afer}\ \emph
  {et~al.}(2015{\natexlab{a}})\citenamefont {Sch\"afer}, \citenamefont {Geles},
  \citenamefont {Rost}, \citenamefont {Rohringer}, \citenamefont {Arrigoni},
  \citenamefont {Held}, \citenamefont {Bl\"umer}, \citenamefont {Aichhorn},\
  and\ \citenamefont {Toschi}}]{Schaefer2015-2}%
  \BibitemOpen
  \bibfield  {author} {\bibinfo {author} {\bibfnamefont {T.}~\bibnamefont
  {Sch\"afer}}, \bibinfo {author} {\bibfnamefont {F.}~\bibnamefont {Geles}},
  \bibinfo {author} {\bibfnamefont {D.}~\bibnamefont {Rost}}, \bibinfo {author}
  {\bibfnamefont {G.}~\bibnamefont {Rohringer}}, \bibinfo {author}
  {\bibfnamefont {E.}~\bibnamefont {Arrigoni}}, \bibinfo {author}
  {\bibfnamefont {K.}~\bibnamefont {Held}}, \bibinfo {author} {\bibfnamefont
  {N.}~\bibnamefont {Bl\"umer}}, \bibinfo {author} {\bibfnamefont
  {M.}~\bibnamefont {Aichhorn}}, \ and\ \bibinfo {author} {\bibfnamefont
  {A.}~\bibnamefont {Toschi}},\ }\href {\doibase 10.1103/PhysRevB.91.125109}
  {\bibfield  {journal} {\bibinfo  {journal} {Phys. Rev. B}\ }\textbf {\bibinfo
  {volume} {91}},\ \bibinfo {pages} {125109} (\bibinfo {year}
  {2015}{\natexlab{a}})}\BibitemShut {NoStop}%
\bibitem [{\citenamefont {Gunnarsson}\ \emph {et~al.}(2015)\citenamefont
  {Gunnarsson}, \citenamefont {Sch\"afer}, \citenamefont {LeBlanc},
  \citenamefont {Gull}, \citenamefont {Merino}, \citenamefont {Sangiovanni},
  \citenamefont {Rohringer},\ and\ \citenamefont {Toschi}}]{Gunnarsson2015}%
  \BibitemOpen
  \bibfield  {author} {\bibinfo {author} {\bibfnamefont {O.}~\bibnamefont
  {Gunnarsson}}, \bibinfo {author} {\bibfnamefont {T.}~\bibnamefont
  {Sch\"afer}}, \bibinfo {author} {\bibfnamefont {J.~P.~F.}\ \bibnamefont
  {LeBlanc}}, \bibinfo {author} {\bibfnamefont {E.}~\bibnamefont {Gull}},
  \bibinfo {author} {\bibfnamefont {J.}~\bibnamefont {Merino}}, \bibinfo
  {author} {\bibfnamefont {G.}~\bibnamefont {Sangiovanni}}, \bibinfo {author}
  {\bibfnamefont {G.}~\bibnamefont {Rohringer}}, \ and\ \bibinfo {author}
  {\bibfnamefont {A.}~\bibnamefont {Toschi}},\ }\href {\doibase
  10.1103/PhysRevLett.114.236402} {\bibfield  {journal} {\bibinfo  {journal}
  {Phys. Rev. Lett.}\ }\textbf {\bibinfo {volume} {114}},\ \bibinfo {pages}
  {236402} (\bibinfo {year} {2015})}\BibitemShut {NoStop}%
\bibitem [{\citenamefont {Gukelberger}\ \emph {et~al.}(2017)\citenamefont
  {Gukelberger}, \citenamefont {Kozik},\ and\ \citenamefont
  {Hafermann}}]{Gukelberger2016}%
  \BibitemOpen
  \bibfield  {author} {\bibinfo {author} {\bibfnamefont {J.}~\bibnamefont
  {Gukelberger}}, \bibinfo {author} {\bibfnamefont {E.}~\bibnamefont {Kozik}},
  \ and\ \bibinfo {author} {\bibfnamefont {H.}~\bibnamefont {Hafermann}},\
  }\href {\doibase 10.1103/PhysRevB.96.035152} {\bibfield  {journal} {\bibinfo
  {journal} {Phys. Rev. B}\ }\textbf {\bibinfo {volume} {96}},\ \bibinfo
  {pages} {035152} (\bibinfo {year} {2017})}\BibitemShut {NoStop}%
\bibitem [{\citenamefont {Sch\"afer}\ \emph {et~al.}(2021)\citenamefont
  {Sch\"afer}, \citenamefont {Wentzell}, \citenamefont {\ifmmode~\check{S}\else
  \v{S}\fi{}imkovic}, \citenamefont {He}, \citenamefont {Hille}, \citenamefont
  {Klett}, \citenamefont {Eckhardt}, \citenamefont {Arzhang}, \citenamefont
  {Harkov}, \citenamefont {Le~R\'egent}, \citenamefont {Kirsch}, \citenamefont
  {Wang}, \citenamefont {Kim}, \citenamefont {Kozik}, \citenamefont {Stepanov},
  \citenamefont {Kauch}, \citenamefont {Andergassen}, \citenamefont {Hansmann},
  \citenamefont {Rohe}, \citenamefont {Vilk}, \citenamefont {LeBlanc},
  \citenamefont {Zhang}, \citenamefont {Tremblay}, \citenamefont {Ferrero},
  \citenamefont {Parcollet},\ and\ \citenamefont {Georges}}]{Schaefer2020}%
  \BibitemOpen
  \bibfield  {author} {\bibinfo {author} {\bibfnamefont {T.}~\bibnamefont
  {Sch\"afer}}, \bibinfo {author} {\bibfnamefont {N.}~\bibnamefont {Wentzell}},
  \bibinfo {author} {\bibfnamefont {F.}~\bibnamefont {\ifmmode~\check{S}\else
  \v{S}\fi{}imkovic}}, \bibinfo {author} {\bibfnamefont {Y.-Y.}\ \bibnamefont
  {He}}, \bibinfo {author} {\bibfnamefont {C.}~\bibnamefont {Hille}}, \bibinfo
  {author} {\bibfnamefont {M.}~\bibnamefont {Klett}}, \bibinfo {author}
  {\bibfnamefont {C.~J.}\ \bibnamefont {Eckhardt}}, \bibinfo {author}
  {\bibfnamefont {B.}~\bibnamefont {Arzhang}}, \bibinfo {author} {\bibfnamefont
  {V.}~\bibnamefont {Harkov}}, \bibinfo {author} {\bibfnamefont {F.~m. c.-M.}\
  \bibnamefont {Le~R\'egent}}, \bibinfo {author} {\bibfnamefont
  {A.}~\bibnamefont {Kirsch}}, \bibinfo {author} {\bibfnamefont
  {Y.}~\bibnamefont {Wang}}, \bibinfo {author} {\bibfnamefont {A.~J.}\
  \bibnamefont {Kim}}, \bibinfo {author} {\bibfnamefont {E.}~\bibnamefont
  {Kozik}}, \bibinfo {author} {\bibfnamefont {E.~A.}\ \bibnamefont {Stepanov}},
  \bibinfo {author} {\bibfnamefont {A.}~\bibnamefont {Kauch}}, \bibinfo
  {author} {\bibfnamefont {S.}~\bibnamefont {Andergassen}}, \bibinfo {author}
  {\bibfnamefont {P.}~\bibnamefont {Hansmann}}, \bibinfo {author}
  {\bibfnamefont {D.}~\bibnamefont {Rohe}}, \bibinfo {author} {\bibfnamefont
  {Y.~M.}\ \bibnamefont {Vilk}}, \bibinfo {author} {\bibfnamefont {J.~P.~F.}\
  \bibnamefont {LeBlanc}}, \bibinfo {author} {\bibfnamefont {S.}~\bibnamefont
  {Zhang}}, \bibinfo {author} {\bibfnamefont {A.-M.~S.}\ \bibnamefont
  {Tremblay}}, \bibinfo {author} {\bibfnamefont {M.}~\bibnamefont {Ferrero}},
  \bibinfo {author} {\bibfnamefont {O.}~\bibnamefont {Parcollet}}, \ and\
  \bibinfo {author} {\bibfnamefont {A.}~\bibnamefont {Georges}},\ }\href
  {\doibase 10.1103/PhysRevX.11.011058} {\bibfield  {journal} {\bibinfo
  {journal} {Phys. Rev. X}\ }\textbf {\bibinfo {volume} {11}},\ \bibinfo
  {pages} {011058} (\bibinfo {year} {2021})}\BibitemShut {NoStop}%
\bibitem [{\citenamefont {Antipov}\ \emph {et~al.}(2014)\citenamefont
  {Antipov}, \citenamefont {Gull},\ and\ \citenamefont
  {Kirchner}}]{Antipov2014}%
  \BibitemOpen
  \bibfield  {author} {\bibinfo {author} {\bibfnamefont {A.~E.}\ \bibnamefont
  {Antipov}}, \bibinfo {author} {\bibfnamefont {E.}~\bibnamefont {Gull}}, \
  and\ \bibinfo {author} {\bibfnamefont {S.}~\bibnamefont {Kirchner}},\
  }\href@noop {} {\bibfield  {journal} {\bibinfo  {journal} {Phys. Rev. Lett.}\
  }\textbf {\bibinfo {volume} {112}},\ \bibinfo {pages} {226401} (\bibinfo
  {year} {2014})}\BibitemShut {NoStop}%
\bibitem [{\citenamefont {Sch\"afer}\ \emph {et~al.}(2017)\citenamefont
  {Sch\"afer}, \citenamefont {Katanin}, \citenamefont {Held},\ and\
  \citenamefont {Toschi}}]{Schaefer2017}%
  \BibitemOpen
  \bibfield  {author} {\bibinfo {author} {\bibfnamefont {T.}~\bibnamefont
  {Sch\"afer}}, \bibinfo {author} {\bibfnamefont {A.~A.}\ \bibnamefont
  {Katanin}}, \bibinfo {author} {\bibfnamefont {K.}~\bibnamefont {Held}}, \
  and\ \bibinfo {author} {\bibfnamefont {A.}~\bibnamefont {Toschi}},\ }\href
  {\doibase 10.1103/PhysRevLett.119.046402} {\bibfield  {journal} {\bibinfo
  {journal} {Phys. Rev. Lett.}\ }\textbf {\bibinfo {volume} {119}},\ \bibinfo
  {pages} {046402} (\bibinfo {year} {2017})}\BibitemShut {NoStop}%
\bibitem [{\citenamefont {Sch\"afer}\ \emph {et~al.}(2019)\citenamefont
  {Sch\"afer}, \citenamefont {Katanin}, \citenamefont {Kitatani}, \citenamefont
  {Toschi},\ and\ \citenamefont {Held}}]{Schaefer2019}%
  \BibitemOpen
  \bibfield  {author} {\bibinfo {author} {\bibfnamefont {T.}~\bibnamefont
  {Sch\"afer}}, \bibinfo {author} {\bibfnamefont {A.~A.}\ \bibnamefont
  {Katanin}}, \bibinfo {author} {\bibfnamefont {M.}~\bibnamefont {Kitatani}},
  \bibinfo {author} {\bibfnamefont {A.}~\bibnamefont {Toschi}}, \ and\ \bibinfo
  {author} {\bibfnamefont {K.}~\bibnamefont {Held}},\ }\href {\doibase
  10.1103/PhysRevLett.122.227201} {\bibfield  {journal} {\bibinfo  {journal}
  {Phys. Rev. Lett.}\ }\textbf {\bibinfo {volume} {122}},\ \bibinfo {pages}
  {227201} (\bibinfo {year} {2019})}\BibitemShut {NoStop}%
\bibitem [{\citenamefont {Werner}\ \emph {et~al.}(2006)\citenamefont {Werner},
  \citenamefont {Comanac}, \citenamefont {de' Medici}, \citenamefont {Troyer},\
  and\ \citenamefont {Millis}}]{Werner2006}%
  \BibitemOpen
  \bibfield  {author} {\bibinfo {author} {\bibfnamefont {P.}~\bibnamefont
  {Werner}}, \bibinfo {author} {\bibfnamefont {A.}~\bibnamefont {Comanac}},
  \bibinfo {author} {\bibfnamefont {L.}~\bibnamefont {de' Medici}}, \bibinfo
  {author} {\bibfnamefont {M.}~\bibnamefont {Troyer}}, \ and\ \bibinfo {author}
  {\bibfnamefont {A.~J.}\ \bibnamefont {Millis}},\ }\href {\doibase
  10.1103/PhysRevLett.97.076405} {\bibfield  {journal} {\bibinfo  {journal}
  {Phys. Rev. Lett.}\ }\textbf {\bibinfo {volume} {97}},\ \bibinfo {pages}
  {076405} (\bibinfo {year} {2006})}\BibitemShut {NoStop}%
\bibitem [{\citenamefont {Gull}\ \emph {et~al.}(2011)\citenamefont {Gull},
  \citenamefont {Millis}, \citenamefont {Lichtenstein}, \citenamefont
  {Rubtsov}, \citenamefont {Troyer},\ and\ \citenamefont {Werner}}]{Gull2011a}%
  \BibitemOpen
  \bibfield  {author} {\bibinfo {author} {\bibfnamefont {E.}~\bibnamefont
  {Gull}}, \bibinfo {author} {\bibfnamefont {A.~J.}\ \bibnamefont {Millis}},
  \bibinfo {author} {\bibfnamefont {A.~I.}\ \bibnamefont {Lichtenstein}},
  \bibinfo {author} {\bibfnamefont {A.~N.}\ \bibnamefont {Rubtsov}}, \bibinfo
  {author} {\bibfnamefont {M.}~\bibnamefont {Troyer}}, \ and\ \bibinfo {author}
  {\bibfnamefont {P.}~\bibnamefont {Werner}},\ }\href {\doibase
  10.1103/RevModPhys.83.349} {\bibfield  {journal} {\bibinfo  {journal} {Rev.
  Mod. Phys.}\ }\textbf {\bibinfo {volume} {83}},\ \bibinfo {pages} {349}
  (\bibinfo {year} {2011})}\BibitemShut {NoStop}%
\bibitem [{\citenamefont {Wallerberger}\ \emph {et~al.}(2019)\citenamefont
  {Wallerberger}, \citenamefont {Hausoel}, \citenamefont {Gunacker},
  \citenamefont {Kowalski}, \citenamefont {Parragh}, \citenamefont {Goth},
  \citenamefont {Held},\ and\ \citenamefont {Sangiovanni}}]{w2dynamics}%
  \BibitemOpen
  \bibfield  {author} {\bibinfo {author} {\bibfnamefont {M.}~\bibnamefont
  {Wallerberger}}, \bibinfo {author} {\bibfnamefont {A.}~\bibnamefont
  {Hausoel}}, \bibinfo {author} {\bibfnamefont {P.}~\bibnamefont {Gunacker}},
  \bibinfo {author} {\bibfnamefont {A.}~\bibnamefont {Kowalski}}, \bibinfo
  {author} {\bibfnamefont {N.}~\bibnamefont {Parragh}}, \bibinfo {author}
  {\bibfnamefont {F.}~\bibnamefont {Goth}}, \bibinfo {author} {\bibfnamefont
  {K.}~\bibnamefont {Held}}, \ and\ \bibinfo {author} {\bibfnamefont
  {G.}~\bibnamefont {Sangiovanni}},\ }\href {\doibase
  https://doi.org/10.1016/j.cpc.2018.09.007} {\bibfield  {journal} {\bibinfo
  {journal} {Computer Physics Communications}\ }\textbf {\bibinfo {volume}
  {235}},\ \bibinfo {pages} {388 } (\bibinfo {year} {2019})}\BibitemShut
  {NoStop}%
\bibitem [{\citenamefont {Gunacker}\ \emph {et~al.}(2015)\citenamefont
  {Gunacker}, \citenamefont {Wallerberger}, \citenamefont {Gull}, \citenamefont
  {Hausoel}, \citenamefont {Sangiovanni},\ and\ \citenamefont
  {Held}}]{Gunacker15}%
  \BibitemOpen
  \bibfield  {author} {\bibinfo {author} {\bibfnamefont {P.}~\bibnamefont
  {Gunacker}}, \bibinfo {author} {\bibfnamefont {M.}~\bibnamefont
  {Wallerberger}}, \bibinfo {author} {\bibfnamefont {E.}~\bibnamefont {Gull}},
  \bibinfo {author} {\bibfnamefont {A.}~\bibnamefont {Hausoel}}, \bibinfo
  {author} {\bibfnamefont {G.}~\bibnamefont {Sangiovanni}}, \ and\ \bibinfo
  {author} {\bibfnamefont {K.}~\bibnamefont {Held}},\ }\href {\doibase
  10.1103/PhysRevB.92.155102} {\bibfield  {journal} {\bibinfo  {journal} {Phys.
  Rev. B}\ }\textbf {\bibinfo {volume} {92}},\ \bibinfo {pages} {155102}
  (\bibinfo {year} {2015})}\BibitemShut {NoStop}%
\bibitem [{\citenamefont {Kaufmann}\ \emph {et~al.}(2021)\citenamefont
  {Kaufmann}, \citenamefont {Eckhardt}, \citenamefont {Pickem}, \citenamefont
  {Kitatani}, \citenamefont {Kauch},\ and\ \citenamefont
  {Held}}]{PhysRevB.103.035120}%
  \BibitemOpen
  \bibfield  {author} {\bibinfo {author} {\bibfnamefont {J.}~\bibnamefont
  {Kaufmann}}, \bibinfo {author} {\bibfnamefont {C.}~\bibnamefont {Eckhardt}},
  \bibinfo {author} {\bibfnamefont {M.}~\bibnamefont {Pickem}}, \bibinfo
  {author} {\bibfnamefont {M.}~\bibnamefont {Kitatani}}, \bibinfo {author}
  {\bibfnamefont {A.}~\bibnamefont {Kauch}}, \ and\ \bibinfo {author}
  {\bibfnamefont {K.}~\bibnamefont {Held}},\ }\href {\doibase
  10.1103/PhysRevB.103.035120} {\bibfield  {journal} {\bibinfo  {journal}
  {Phys. Rev. B}\ }\textbf {\bibinfo {volume} {103}},\ \bibinfo {pages}
  {035120} (\bibinfo {year} {2021})}\BibitemShut {NoStop}%
\bibitem [{Note1()}]{Note1}%
  \BibitemOpen
  \bibinfo {note} {Due to the multi-orbital nature and the temperature scaling
  of the D$\Gamma $A we are restricted in temperature, i.e. we are not able to
  approach the FM instability much further.}\BibitemShut {Stop}%
\bibitem [{\citenamefont {Gunnarsson}\ \emph {et~al.}(1996)\citenamefont
  {Gunnarsson}, \citenamefont {Koch},\ and\ \citenamefont
  {Martin}}]{PhysRevB.54.R11026}%
  \BibitemOpen
  \bibfield  {author} {\bibinfo {author} {\bibfnamefont {O.}~\bibnamefont
  {Gunnarsson}}, \bibinfo {author} {\bibfnamefont {E.}~\bibnamefont {Koch}}, \
  and\ \bibinfo {author} {\bibfnamefont {R.~M.}\ \bibnamefont {Martin}},\
  }\href {\doibase 10.1103/PhysRevB.54.R11026} {\bibfield  {journal} {\bibinfo
  {journal} {Phys. Rev. B}\ }\textbf {\bibinfo {volume} {54}},\ \bibinfo
  {pages} {R11026} (\bibinfo {year} {1996})}\BibitemShut {NoStop}%
\bibitem [{\citenamefont {Pavarini}\ \emph {et~al.}(2004)\citenamefont
  {Pavarini}, \citenamefont {Biermann}, \citenamefont {Poteryaev},
  \citenamefont {Lichtenstein}, \citenamefont {Georges},\ and\ \citenamefont
  {Andersen}}]{pavarini:176403}%
  \BibitemOpen
  \bibfield  {author} {\bibinfo {author} {\bibfnamefont {E.}~\bibnamefont
  {Pavarini}}, \bibinfo {author} {\bibfnamefont {S.}~\bibnamefont {Biermann}},
  \bibinfo {author} {\bibfnamefont {A.}~\bibnamefont {Poteryaev}}, \bibinfo
  {author} {\bibfnamefont {A.~I.}\ \bibnamefont {Lichtenstein}}, \bibinfo
  {author} {\bibfnamefont {A.}~\bibnamefont {Georges}}, \ and\ \bibinfo
  {author} {\bibfnamefont {O.~K.}\ \bibnamefont {Andersen}},\ }\href
  {http://link.aps.org/abstract/PRL/v92/e176403} {\bibfield  {journal}
  {\bibinfo  {journal} {Phys. Rev. Lett.}\ }\textbf {\bibinfo {volume} {92}},\
  \bibinfo {eid} {176403} (\bibinfo {year} {2004})}\BibitemShut {NoStop}%
\bibitem [{\citenamefont {Pavarini}\ \emph {et~al.}(2001)\citenamefont
  {Pavarini}, \citenamefont {Dasgupta}, \citenamefont {Saha-Dasgupta},
  \citenamefont {Jepsen},\ and\ \citenamefont
  {Andersen}}]{PhysRevLett.87.047003}%
  \BibitemOpen
  \bibfield  {author} {\bibinfo {author} {\bibfnamefont {E.}~\bibnamefont
  {Pavarini}}, \bibinfo {author} {\bibfnamefont {I.}~\bibnamefont {Dasgupta}},
  \bibinfo {author} {\bibfnamefont {T.}~\bibnamefont {Saha-Dasgupta}}, \bibinfo
  {author} {\bibfnamefont {O.}~\bibnamefont {Jepsen}}, \ and\ \bibinfo {author}
  {\bibfnamefont {O.~K.}\ \bibnamefont {Andersen}},\ }\href {\doibase
  10.1103/PhysRevLett.87.047003} {\bibfield  {journal} {\bibinfo  {journal}
  {Phys. Rev. Lett.}\ }\textbf {\bibinfo {volume} {87}},\ \bibinfo {pages}
  {047003} (\bibinfo {year} {2001})}\BibitemShut {NoStop}%
\bibitem [{\citenamefont {Held}\ and\ \citenamefont
  {Vollhardt}(1998)}]{Held1998}%
  \BibitemOpen
  \bibfield  {author} {\bibinfo {author} {\bibfnamefont {K.}~\bibnamefont
  {Held}}\ and\ \bibinfo {author} {\bibfnamefont {D.}~\bibnamefont
  {Vollhardt}},\ }\href {\doibase 10.1007/s100510050468} {\bibfield  {journal}
  {\bibinfo  {journal} {The European Physical Journal B - Condensed Matter and
  Complex Systems}\ }\textbf {\bibinfo {volume} {5}},\ \bibinfo {pages} {473}
  (\bibinfo {year} {1998})}\BibitemShut {NoStop}%
\bibitem [{\citenamefont {Sch\"afer}\ \emph
  {et~al.}(2015{\natexlab{b}})\citenamefont {Sch\"afer}, \citenamefont
  {Toschi},\ and\ \citenamefont {Tomczak}}]{jmt_dga3d}%
  \BibitemOpen
  \bibfield  {author} {\bibinfo {author} {\bibfnamefont {T.}~\bibnamefont
  {Sch\"afer}}, \bibinfo {author} {\bibfnamefont {A.}~\bibnamefont {Toschi}}, \
  and\ \bibinfo {author} {\bibfnamefont {J.~M.}\ \bibnamefont {Tomczak}},\
  }\href {\doibase 10.1103/PhysRevB.91.121107} {\bibfield  {journal} {\bibinfo
  {journal} {Phys. Rev. B}\ }\textbf {\bibinfo {volume} {91}},\ \bibinfo
  {pages} {121107(R)} (\bibinfo {year} {2015}{\natexlab{b}})}\BibitemShut
  {NoStop}%
\bibitem [{\citenamefont {Mahan}\ and\ \citenamefont
  {Sofo}(1996)}]{Mahan07231996}%
  \BibitemOpen
  \bibfield  {author} {\bibinfo {author} {\bibfnamefont {G.}~\bibnamefont
  {Mahan}}\ and\ \bibinfo {author} {\bibfnamefont {J.}~\bibnamefont {Sofo}},\
  }\href {\doibase 10.1073/pnas.93.15.7436} {\bibfield  {journal} {\bibinfo
  {journal} {Proc. Nat. Acad. Sci. USA}\ }\textbf {\bibinfo {volume} {93}},\
  \bibinfo {pages} {7436} (\bibinfo {year} {1996})}\BibitemShut {NoStop}%
\bibitem [{\citenamefont {Held}\ \emph {et~al.}(2009)\citenamefont {Held},
  \citenamefont {Arita}, \citenamefont {Anisimov},\ and\ \citenamefont
  {Kuroki}}]{karsten_hvar}%
  \BibitemOpen
  \bibfield  {author} {\bibinfo {author} {\bibfnamefont {K.}~\bibnamefont
  {Held}}, \bibinfo {author} {\bibfnamefont {R.}~\bibnamefont {Arita}},
  \bibinfo {author} {\bibfnamefont {V.~I.}\ \bibnamefont {Anisimov}}, \ and\
  \bibinfo {author} {\bibfnamefont {K.}~\bibnamefont {Kuroki}},\ }in\ \href
  {https://link.springer.com/chapter/10.1007/978-90-481-2892-1_9} {\emph
  {\bibinfo {booktitle} {Properties and Applications of Thermoelectric
  Materials}}},\ \bibinfo {series and number} {NATO Science for Peace and
  Security Series B: Physics and Biophysics},\ \bibinfo {editor} {edited by\
  \bibinfo {editor} {\bibfnamefont {V.}~\bibnamefont {Zlati{\'c}}}\ and\
  \bibinfo {editor} {\bibfnamefont {A.}~\bibnamefont {Hewson}}}\ (\bibinfo
  {publisher} {Springer},\ \bibinfo {year} {2009})\ p.\ \bibinfo {pages}
  {141}\BibitemShut {NoStop}%
\bibitem [{\citenamefont {Zlatic}\ and\ \citenamefont
  {Monnier}(2014)}]{zlatic}%
  \BibitemOpen
  \bibfield  {author} {\bibinfo {author} {\bibfnamefont {V.}~\bibnamefont
  {Zlatic}}\ and\ \bibinfo {author} {\bibfnamefont {R.}~\bibnamefont
  {Monnier}},\ }\href@noop {} {\emph {\bibinfo {title} {Modern Theory of
  Thermoelectricity}}}\ (\bibinfo  {publisher} {Oxford University Press},\
  \bibinfo {year} {2014})\BibitemShut {NoStop}%
\bibitem [{\citenamefont {Tomczak}(2018)}]{NGCS}%
  \BibitemOpen
  \bibfield  {author} {\bibinfo {author} {\bibfnamefont {J.~M.}\ \bibnamefont
  {Tomczak}},\ }\href {http://stacks.iop.org/0953-8984/30/i=18/a=183001}
  {\bibfield  {journal} {\bibinfo  {journal} {J. Phys.: Condens. Matter
  (Topical Review)}\ }\textbf {\bibinfo {volume} {30}},\ \bibinfo {pages}
  {183001} (\bibinfo {year} {2018})}\BibitemShut {NoStop}%
\bibitem [{\citenamefont {Haule}\ and\ \citenamefont
  {Kotliar}(2009)}]{10.1007/978-90-481-2892-1_7}%
  \BibitemOpen
  \bibfield  {author} {\bibinfo {author} {\bibfnamefont {K.}~\bibnamefont
  {Haule}}\ and\ \bibinfo {author} {\bibfnamefont {G.}~\bibnamefont
  {Kotliar}},\ }in\ \href@noop {} {\emph {\bibinfo {booktitle} {Properties and
  Applications of Thermoelectric Materials}}},\ \bibinfo {editor} {edited by\
  \bibinfo {editor} {\bibfnamefont {V.}~\bibnamefont {Zlati{\'{c}}}}\ and\
  \bibinfo {editor} {\bibfnamefont {A.~C.}\ \bibnamefont {Hewson}}}\ (\bibinfo
  {publisher} {Springer Netherlands},\ \bibinfo {address} {Dordrecht},\
  \bibinfo {year} {2009})\ pp.\ \bibinfo {pages} {119--131}\BibitemShut
  {NoStop}%
\bibitem [{\citenamefont {Sun}\ \emph {et~al.}(2013)\citenamefont {Sun},
  \citenamefont {Xu}, \citenamefont {Tomczak}, \citenamefont {Kotliar},
  \citenamefont {S\o{}ndergaard}, \citenamefont {Iversen},\ and\ \citenamefont
  {Steglich}}]{PhysRevB.88.245203}%
  \BibitemOpen
  \bibfield  {author} {\bibinfo {author} {\bibfnamefont {P.}~\bibnamefont
  {Sun}}, \bibinfo {author} {\bibfnamefont {W.}~\bibnamefont {Xu}}, \bibinfo
  {author} {\bibfnamefont {J.~M.}\ \bibnamefont {Tomczak}}, \bibinfo {author}
  {\bibfnamefont {G.}~\bibnamefont {Kotliar}}, \bibinfo {author} {\bibfnamefont
  {M.}~\bibnamefont {S\o{}ndergaard}}, \bibinfo {author} {\bibfnamefont
  {B.~B.}\ \bibnamefont {Iversen}}, \ and\ \bibinfo {author} {\bibfnamefont
  {F.}~\bibnamefont {Steglich}},\ }\href {\doibase 10.1103/PhysRevB.88.245203}
  {\bibfield  {journal} {\bibinfo  {journal} {Phys. Rev. B}\ }\textbf {\bibinfo
  {volume} {88}},\ \bibinfo {pages} {245203} (\bibinfo {year}
  {2013})}\BibitemShut {NoStop}%
\end{thebibliography}
\end{document}